\documentclass[12pt]{article}
\usepackage{amsmath}
\usepackage{amssymb}
\begin{document}
%\begin{flushright}
%BHU-PHYS-CAS Preprint\\
%arXiv: 1005.5067 [hep-th]
%\end{flushright}
\begin{center}
{\bf {\large{1D Diffeomorphism Invariant  Model of a Free Scalar Relativistic Particle: Supervariable and BRST Approaches}}}

\vskip 2.5cm

{\sf  B. Chauhan$^{(a)}$, A. K. Rao $^{(a)}$, A. Tripathi $^{(a)}$,  R. P. Malik$^{(a,b)}$}\\
$^{(a)}$ {\it Physics Department, Institute of Science,}\\
{\it Banaras Hindu University, Varanasi - 221 005, (U.P.), India}\\

\vskip 0.1cm

\vskip 0.1cm

$^{(b)}$ {\it DST Centre for Interdisciplinary Mathematical Sciences,}\\
{\it Institute of Science, Banaras Hindu University, Varanasi - 221 005, India}\\
{\small {\sf {e-mails: bchauhan501@gmail.com; amit.akrao@gmail.com;\\ ankur1793@gmail.com; rpmalik1995@gmail.com}}}
\end{center}

\vskip 1.3 cm

\noindent
{\bf Abstract:} We apply the supervariable approach to derive the proper  
{\it quantum} Becchi-Rouet-Stora-Tyutin (BRST) and anti-BRST symmetries for
 the 1D diffeomorphism invariant  model of a free scalar relativistic particle  by exploiting the infinitesimal  
{\it classical} reparameterization (i.e. 1D diffeomorphism) symmetry of {\it this} theory. 
 We derive the conserved and off-shell nilpotent (anti-)BRST charges and prove their absolute
 anticommutativity property by using the virtues of  Curci-Ferrari (CF)-type restriction of our present theory. 
We establish the sanctity of the existence of CF-type restriction (i) by considering the (anti-)BRST
 symmetry transformations of  the coupled (but equivalent) Lagrangians, and (ii) by proving the symmetry invariance of the 
Lagrangians within the framework of supervariable approach. We capture the nilpotency and
absolute anticommutativity of the conserved (anti-)BRST charges within the framework of (anti-)chiral
supervariable approach (ACSA) to BRST formalism. One of the novel observations of our present endeavor
is the derivation of  CF-type restriction by using the {\it modified} Bonora-Tonin (BT) supervariable
approach (while deriving the (anti-)BRST symmetries for the {\it target} spacetime and/or momenta variables)
{\it and} by symmetry considerations of the Lagrangians of the theory. 
The rest of the (anti-)BRST  symmetries, for the other
variables, are derived by using the {\it newly} proposed ACSA. 
We also demonstrate the existence of CF-type restriction in the proof of absolute anticommutativity 
of the (anti-)BRST charges.



\newpage

\section{Introduction}

\noindent
For the covariant canonical quantization of the gauge theories \big [characterized by the first-class constraints in the 
terminology of  Dirac's prescription for the  classification scheme of constraints \big (see, e.g. [1, 2]\big )\big ], one of 
the most intuitive,  instructive and mathematically rich approaches is the Becchi-Rouet-Stora-Tyutin (BRST)
 formalism [3-6] which is also 
useful in the quantization of the diffeomorphism invariant theories. Two of the key characteristic
 features of the BRST formalism are the nilpotency and absolute anticommutativity properties associated with the (anti-) BRST symmetries
 which exist at the {\it quantum} level corresponding to an infinitesimal {\it classical} local gauge (and/or diffeomorphism)
 symmetry transformation.  The geometrical superfield/supervariable approach [7-14] to BRST formalism provides the geometrical origin and 
interpretation for the above cited {\it two} key properties that are associated with the {\it quantum gauge} \big[i.e. (anti-) BRST\big] symmetries. 
%The beauty of the BRST formalism is the observation that the {\it unitarity} and {\it quantum} gauge (i.e. BRST) invariance are respected 
%{\it together}  at any arbitrary order of perturbative computations for a given process (allowed by the {\it interacting} gauge theory).       

The usual superfield approach (USFA) to BRST formalism [7-11] exploits the idea of horizontality condition (HC) where a 
$(p + 1)$-form curvature tensor (i.e. field strength tensor), corresponding to a given $p$-form ($p = 1, 2, 3, ...$)
 gauge field, plays a pivotal role [9-11]. In fact, the application of the USFA leads to the precise 
derivation of (i) the off-shell nilpotent  and absolutely anticommuting  (anti-)BRST symmetry transformations for the $p$-form
 {\it gauge} and associated {\it (anti-)ghost} fields for a given $p$-form gauge theory, and (ii) the (anti-)BRST invariant  Curci-Ferrari (CF)-condition  [9]. It does not shed any light, however, on the derivation of the (anti-)BRST symmetries for the {\it matter} fields in an {\it interacting} $p$-form gauge theory. 
The USFA has been systematically generalized so as to derive the (anti-)BRST symmetries for the gauge, (anti-)ghost and
 {\it matter} fields {\it together} by invoking the gauge invariant restrictions (GIRs) in addition to the HC. This extended version 
(see, e.g. [12-14]) of the USFA has been christened as the augmented version of superfield/supervariable approach (AVSA) to BRST 
formalism. It may be mentioned here that {\it both} the  HC and GIRs are found to be {\it consistent} with each-other and, primarily, 
they {\it complement} each-other beautifully.

All the above developments have been achieved in the context of $p$-form gauge theories {\it only}. It has been a long-standing
 problem to apply the AVSA/USFA to the {\it diffeomorphism} invariant theories which are very important in
 the context of modern developments in the gravitational and (super)string theories. Against this backdrop, it is 
pertinent to point out that we have applied the  (anti-)chiral superfield/supervariable approach 
(ACSA) (see, e.g. [15-18]) to derive the (anti-)BRST symmetries for {\it all} the relevant variables of a set of two {\it reparameterization}
 invariant theories [18]. These theories are the  models [i.e. (0 + 1)-dimensional (1D) models] of a free {\it scalar}
 as well as  a {\it spinning} relativistic
 particle. However, these theories are {\it also} found to be invariant under the {\it gauge} symmetry transformations.
The latter symmetry transformations are found to be {\it equivalent} to the reparameterization symmetries in some specific limits
(see, e.g. [19, 18] for details) where the on-shell condition {\it and} some choices are made in an {\it ad-hoc} fashion. Thus, the models considered in [18] are 
{\it not} purely reparameterization (i.e. 1D diffeomorphism) invariant theories. The {\it classical} gauge symmetries of these theories have been 
exploited in the context of their BRST quantization (see, e.g. [19] for details).

In a quite recent set of works\footnote{We christen the geometrical BT-superfield/supervariable approach [9-11] 
 as the {\it modified} BT-superfield/supervariable approach (MBTSA) to BRST  formalism 
[20, 21] when  we take into account the {\it ordinary} infinitesimal {\it diffeomorphism/reparameterization}
  transformation [cf. Eq. (5) below] and its generalization to  the (1, 2)-dimensional superspace infinitesimal
  diffeomorphism/reparameterization transformation [cf. Eq. (13) below] that is defined on 
a (1, 2)-dimensional supermanifold.} in Refs. [20, 21], it has been shown that
 the Bonora-Tonin (BT) superfield formalism [9-11] can be applied to the D-dimensional diffeomorphism invariant 
theories {\it provided} we take into account the generalized version of the  infinitesimal diffeomorphism 
transformations [cf. Eq. (13) below] in the {\it superfields/supervariables}  that are 
 defined on the (D, 2)-dimensional supermanifold. The {\it latter} is parametrized by the superspace coordinates $Z^M \equiv (x^\mu, \theta, \bar{\theta})$
 where $x^\mu (\mu = 0, 1, 2, ...D-1)$ are the bosonic variables and $(\theta, \bar{\theta})$ are a pair of Grassmannian variables 
that satisfy: $\theta^2 = \bar{\theta}^2 = 0, \,\,\,\theta\,\bar{\theta} + \bar{\theta}\,\theta = 0$. We perform the {\it proper} super expansions of the 
supervariables/superfields along the $\theta$ and $\bar{\theta}$-directions of the 
(D, 2)-dimensional  supermanifold. The restrictions that are imposed on the superfields/supervariables have been called as the HC because 
the (super)exterior derivatives play very important roles in these restrictions. The importance of the (super)exterior derivatives becomes very clear when one 
derives the (anti-)BRST symmetries (corresponding to the infinitesimal diffeomorphism transformation) for the {\it vector}
 and {\it metric} tensors of the theory [20, 21]. In this analysis, only scalar superfield/supervariable is  an {\it exception}
 where there happens to be {\it no} use of the (super)exterior derivatives in any kind of restrictions. Despite this, the simple and straightforward
restriction that is imposed on the supervariable/superfield is still called as the HC (see, e.g. [20, 21]). In our present endeavor, we deal {\it only} 
with the {\it scalar} (super)variables that are defined on the (1, 2)-dimensional supermanifold.

We have applied, in our present endeavor,  the {\it modified} BT-supervariable approach (MBTSA) to BRST formalism proposed in 
Refs. [20, 21] to a 1D diffeomorphism
 (i.e. reparameterization) invariant  theory of a free {\it scalar} relativistic particle and obtained the proper 
(i.e. off-shell nilpotent and absolutely anticommuting) (anti-)BRST symmetries for the target space variables and 
CF-type restriction for the {\it first} time. The restriction(s) that have been imposed  to derive the (anti-)BRST 
symmetries and CF-type restriction have been called as the HC.  The existence of the 
CF-type restriction is the hallmark of a {\it quantum} theory when the {\it latter} is discussed within the 
framework of BRST formalism [22, 23]. We have applied the ACSA to obtain {\it all} the rest of the (anti-)BRST 
 transformations for the {\it other} variables [i.e. auxiliary and (anti-)ghost variables] of our BRST
invariant  theory. This has led us to derive the appropriate coupled (but equivalent) Lagrangians for our 
theory which {\it individually} respect the (anti-)BRST transformations provided we restrict our discussions
 on a submanifold of the space of quantum  variables (in the {\it total} quantum Hilbert space) where the CF-type
restriction of our theory is satisfied.

The following key factors have been responsible for our curiosity in pursuing our present investigation.
First, the diffeomorphism 
invariance is one of the key features of the gravitational theories in general and superstring theory in particular.
Thus, it is important  to apply the superfield/supervariable  approache to discuss the BRST quantization of such theories. Second,
our present model of the free {\it scalar} relativistic particle 
is a reparameterization invariant theory whose  generalization is nothing but the bosonic string theory [24]. 
Thus, it is crucial  to carry out  {\it its} BRST 
analysis and derive the associated CF-type restriction. Third, {\it this} model is interesting in its own right because it is endowed with the 
{\it gauge} as well as {\it reparameterization} symmetries 
which are found to be {\it equivalent} under specific restrictions (cf. Sec. 2 below).
Fourth, the gauge symmetry has been exploited for the BRST quantization in the standard literature 
(see, e.g. [19, 18]). We exploit the infinitesimal {\it classical} reparameterization symmetry 
for the BRST analysis in our present endeavor as, to the best of our knowledge, this has {\it not} been accomplished elsewhere. 
 Finally, our present discussion is our modest first-step towards our central goal to apply the    
superfield  {\it approach} to {\it any} arbitrary D-dimensional  diffeomorphism invariant theory.

The contents of our present endeavor  are organized as follows. In Sec. 2, we discuss the bare essentials of the 
{\it classical} gauge and reparameterization symmetries of the one (0 + 1)-dimensional (1D) model of a {\it scalar} 
relativistic particle and establish {\it their} equivalence. Our Sec. 3  deals with 
the derivation of CF-type restriction in the context  of {\it quantum} nilpotent  (anti-)BRST symmetries (corresponding
to the {\it classical} reparameterization symmetry) within the framework of supervariable formalism where 
the {\it full} super expansions of the supervariables, on the  suitably chosen (1, 2)-dimensional supermanifold, are taken 
into account. We also obtain here the (anti-)BRST symmetry transformations for the target spacetime and momenta 
variables. Our Sec. 4 contains the theoretical material where we derive the {\it rest} of the (anti-)BRST symmetries by using the ACSA [15-18]. Sec. 5 of our paper describes the invariance of the {\it coupled} Lagrangians in the ordinary spacetime.
 Sec. 6 of our present investigation is devoted to capture the
 invariance of the Lagrangians, off-shell nilpotency and absolute anticommutativity of the conserved (anti-)BRST 
charges within the framework of ACSA. Finally, in Sec. 7, we summarize our key results and point
 out a few future directions for further investigation(s).

Our Appendix A is devoted to the proof of the (anti-)BRST invariance of the CF-type restriction 
[cf. Eq. (24) below] of our theory. In our Appendix B, we provide an alternative proof for the existence 
of the  CF-type restriction by proving  the absolute anticommutativity of the conserved and nilpotent (anti-)BRST charges.\\

\section{Preliminaries: Lagrangian Formulation and Some Continuous Symmetry  Transformations}

We begin with the following  {\it three equivalent} Lagrangians for the  free {\it scalar} relativistic particle as
(see, e.g. [19], [18] for details)
\begin{eqnarray}
&& L_0 = m\,\sqrt {\dot x^2},\qquad L_f = p_\mu\,\dot x^\mu - \frac {e}{2}\;(p^2 - m^2),\qquad L_s  = \frac {1}{2\,e}\; \dot x^2 + \frac {e}{2}\;m^2,
\end{eqnarray}
where $L_0$ is the Lagrangian with a square-root, $L_f$ is the {\it first}-order Lagrangian and 
$L_s$ is the {\it second}-order Lagrangian.
The trajectory of the 1D free scalar relativistic particle is parametrized by the 
evolution parameter $\tau$ and $\dot x_\mu = \frac {dx_\mu}{d\tau}$
(with $\mu = 0, 1, 2,...,D-1)$ are the generalized ``velocities" of the free
 particle with momenta $p_\mu$ and rest mass $m$. In the above, Lagrangians
$L_f$ and $L_s$ contain a Lagrangian multiplier variable which is called as the
  einbein variable and it behaves like a 
``gauge" variable in our theory (see, e.g. [19]). It is to be noted that the trajectory of our 1D toy model is embedded in a D-dimensional flat Minkowskian spacetime manifold. The latter acts as the {\it target} spacetime manifold 
[with $x^\mu (\mu = 0, 1, 2, 3...D-1), \; \partial_\mu =  {\partial/ \partial {x^\mu}}$, etc.].

We shall focus, in our present investigation, on the {\it first-order} Lagrangian $L_f$ (because $L_0$ has a
square-root and $L_s$ has a variable in the denominator). This first-order Lagrangian is endowed with the following {\it first-class} 
constraints in the terminology of Dirac's prescription for the classification scheme  [1, 2], namely; 
\begin{eqnarray}
\Pi _{(e)}\approx 0, \qquad\quad -\,\frac {1}{2}\, (p^2 - m^2) \approx  0,
\end{eqnarray}
where $\Pi _{(e)}$ is the canonical conjugate momentum w.r.t. $e$ and $p^2 - m^2 = 0$ is the mass-shell condition.
It is evident that $\Pi _{(e)} \approx 0$ is the {\it primary} constraint and $p^2 - m^2 \approx 0$ 
is the {\it secondary} constraint on the theory. These constraints are 
at the heart of the presence of a {\it gauge} symmetry transformation in the
 theory because the {\it latter} is generated by the following  generator ($G$)
\begin{eqnarray}
G = \dot\xi\, \Pi _{(e)}+ \frac {1}{2}\,\xi\,(p^2 - m^2),
\end{eqnarray} 
where $\xi (\tau)$ is the infinitesimal gauge transformation parameter. It is obvious  that {\it both} the {\it first-class} 
constraints are present in the generator $G$
of the gauge symmetry transformations: $\delta_g\, x_\mu = \xi\,p_\mu, \delta_g\, p_\mu = 0, \delta_g e = \dot\xi$ which are derived from the 
general formula $\delta_g \,\phi  = -\,i\,[\phi, G]$ for the {\it generic} variable $\phi  = x_\mu, p_\mu, e$. In the above derivation, we have
 to use the {\it non-vanishing} standard commutators $[x_\mu, p^\nu] = i\,{\hbar}\,\delta_\mu ^{\nu}$ and $[e, \Pi _{(e)}] = i\,{\hbar}$
 and take the natural units ${\hbar}= c = 1.$ The above gauge symmetry transformations $(\delta_g)$ lead to
 the variation of the first-order Lagrangian $L_f$ as
\begin{eqnarray}
\delta _g\, L_f = \frac {d}{d\tau}\;\Big [\frac {1}{2}\; \xi\, (p^2 + m^2) \Big],
\end{eqnarray}
thereby rendering the action integral $S = \int_{-\,\infty}^{+\, \infty} d\tau\, L_f $ invariant for 
the {\it physically} well-defined parameter $\xi (\tau)$ and the target 
space momenta variables $p_\mu (\tau)$ which vanish-off as $\tau \longrightarrow  \pm\,\infty$ (i.e.  the 
limiting case when   $\tau \longrightarrow \pm \infty$).

The first-order Lagrangian  $L_f$ {\it also} respects an infinitesimal reparameterization symmetry 
$(\delta_r)$ as given below (see, e.g.  [19, 18] for details)
\begin{eqnarray}
&&\delta _r\, x_\mu = \epsilon \,\dot x_\mu, \quad \delta _r\, p_\mu = \epsilon  \,\dot p_\mu,\quad \delta _r\,e = \frac {d}{d\tau}\, (\epsilon  \,e),
\end{eqnarray}
where $\epsilon (\tau) $ is the infinitesimal transformation parameter 
in\footnote{ Actual reparameterization symmetry transformation 
is: $\tau \rightarrow \tau^{\prime} = f(\tau)$ where $f(\tau)$ is a {\it physically}
 well-defined function of $\tau$. However, this function is taken as: 
$f(\tau) = \tau -\epsilon (\tau)$ for {\it its} infinitesimal version where $\epsilon (\tau)$ is the infinitesimal transformation parameter.}
: $\tau\rightarrow \tau - \epsilon (\tau)$. In fact,
under (5), the Lagrangian $L_f$ transforms as: $\delta _r\, L_f = \frac {d}{d\tau}\,(\epsilon  \,L_f)$ 
thereby rendering the action integral $S = \int_{-\,\infty}^{+\, \infty} d\tau\, L_f $ invariant.
A close look at the gauge and reparameterization symmetry transformations demonstrates that {\it both}
these continuous symmetries are {\it equivalent } on-shell (i.e. $\dot x_\mu = e\,p_\mu,\,\,\, \dot p_\mu = 0$)
provided we identify the gauge transformation parameter $\xi $ with the infinitesimal reparameterization transformation
parameter $\epsilon $ as: $\xi = e\,\epsilon$ (where $e$ is the einbein variable that is present in our theory as 
a ``gauge" variable and/or as a Lagrange multiplier).

In literature [19], the {\it classical} gauge symmetry $(\delta_g)$ has been elevated to the
{\it quantum} gauge [i.e. (anti-)BRST] symmetries for our present theory, namely;
 \begin{eqnarray}
&&s_{ab}\, x_\mu = \bar c \;p_\mu, \; \; s_{ab} \,\bar c = 0, \; \; s_{ab}\, p_\mu = 0,  \;\; s_{ab}\, c = - i\, b,\;\; s_{ab} b = 0, \;\; s_{ab}\,\; e = \dot {\bar c},\nonumber\\
&&s_b\, x_\mu = c \, \; p_\mu, \;\;s_b\, c = 0, \;\;\; s_b \,p_\mu = 0, \;\;\; s_b\, \bar c = i\,\; b, \;\;\; s_b \,b = 0, \;\;\; s_b\, \; e = \dot c,  
\end{eqnarray}
which are respected  by a {\it{single}} Lagrangian [19]
\begin{eqnarray}
L_{b} =   p_\mu \; \dot x^\mu - \frac{1}{2}\;e\; (p^2 - m^2) + b\; \dot e 
+ \frac{1}{2}\; b^2 - i\; \dot {\bar c}\; \dot c,
\end{eqnarray}
where $b (\tau)$ is the Nakanishi-Lautrup type {\it bosonic} auxiliary variable, $(\bar c)c$ are the 
{\it fermionic} $(c^2 = \bar c^2 = 0, \, c\, \bar c + \bar c\, c = 0)$ (anti-)ghost
variables and  the gauge-fixing and Faddeev-Popov ghost terms have been 
derived\footnote{The derivation of the gauge-fixing and Faddeev-Popov ghost 
terms [cf. Eq. (8)] is {\it exactly } same as the {\it ones } that are used for the Abelian 1-form ($A^{(1)} = dx^\mu\,A_\mu $) 
Maxwell's $ U(1)$ gauge theory where the gauge field $A_\mu$ 
has been replaced by the ``gauge" variable $e (\tau)$ in our reparameterization (1D diffeomorphism) 
invariant theory for the BRST analysis.} from the following {\it three} explicit variations w.r.t. the 
(anti-) BRST symmetries $s_{(a)b}$, namely; 
\begin{eqnarray}
&&s_b \,s_{ab} \Big [\frac {i}{2}\, e^2 - \frac {\bar c\, c}{2}\Big ],\qquad s_b \,\Big [-\,i\,\bar c \,\Big (\dot e 
+ \frac {b}{2} \Big)\Big ],\qquad s_{ab} \,\Big [\,i\, c \,\Big (\dot e + \frac {b}{2} \Big)\Big ],
\end{eqnarray}
modulo some total derivatives w.r.t. the evolution parameter $\tau$.
It is elementary to check that we have the following explicit (anti-)BRST symmetry  transformations for the Lagrangian $L_f$, namely;
\begin{eqnarray}
&&s_{ab} L_b = {\displaystyle \frac{d} {d \tau}} \;
\Bigl [ \frac{1}{2}\; \bar c\; (p^2 + m^2) + b\; \dot {\bar c} \Big ],\quad s_b L_b = {\displaystyle \frac{d} {d \tau}} \;
\Bigl [ \frac{1}{2}\; c\; (p^2 + m^2) + b\; \dot c \Bigr ], 
\end{eqnarray}
which demonstrate  that the (anti-)BRST symmetries $(6)$ are the symmetries of the action integral
 $S = \int_{-\,\infty}^{+\, \infty}{d\,\tau\,L_b}$ because of the Gauss divergence theorem.

We end this section with the following remarks. First, we observe that there is a {\it{single}} Lagrangian that respects
 {\it both} the BRST as well as the anti-BRST symmetries corresponding to the {\it{classical}} gauge  
transformations: $\delta_g\,x_\mu = \xi\,p_\mu,\, \delta_g\,p_\mu = 0,\, \delta_g\,e = \dot{\xi}$. Second, 
the (anti-)BRST symmetries $s_{(a)b}$ are off-shell nilpotent $(s_{(a)b}^2 = 0)$ and absolutely anticommuting in nature (i.e. $s_b\,s_{ab} + s_{ab}\,s_b = 0$). As a consequence of the above observation, it can be checked that the following is true:
\begin{eqnarray}
s_b \,s_{ab} \Big [\frac {i}{2}\, e^2 - \frac {\bar c\, c}{2}\Big ]\;\;\; \equiv \;\;-\,s_{ab}\, s_{b} \Big [\frac {i}{2}\, e^2 - \frac {\bar c\, c}{2}\Big].
\end{eqnarray}
The above result establishes the fact that the gauge-fixing and Faddeev-Popov ghost terms are (anti-)BRST invariant
 due to the off-shell nilpotency $(s_{(a)b}^2 = 0)$ of the {\it fermionic}  (anti-)BRST symmetry transformations. 
Third, the above {\it{quantum}} (anti-)BRST symmetries are continuous. Thus, the Noether theorem leads to the derivation
 of the following conserved charges as the generators for the (anti-)BRST symmetry transformations (6), namely;  
\begin{eqnarray}
&&Q_{ab}  = \frac {\bar c}{2}\; (p^2-m^2) + b\,\dot {\bar c}\equiv b\,\dot {\bar c}- \dot b\,{\bar c}, \nonumber\\
 &&Q_b  = \frac {c}{2}\; (p^2-m^2) + b\,\dot c\equiv b\,\dot c- \dot b\,c.
\end{eqnarray}  
Fourth, the off-shell nilpotency and absolute anticommutativity of these charges can be proven by using the {\it basic} principle
 behind the continuous symmetry transformations and their generators (as the conserved Noether charges). In 
 other words, we have the following relationships 
\begin{eqnarray}
&&s_b\,Q_b  = -i\, {\{Q_b, Q_b}\} = 0,\nonumber\\
&& s_{ab}\,Q_b = -\, i \;{\{Q_b, Q_{ab}}\} = 0,\nonumber\\
&&s_{ab}\,Q_{ab} = -i\,{\{ Q_{ab},Q_{ab}}\} = 0, \nonumber\\
&&s_b\,Q_{ab} = - \,i\; {\{Q_{ab}, Q_b}\} = 0,
\end{eqnarray} 
where the l.h.s. can be computed {\it easily} by applying {\it directly} the (anti-)BRST symmetry transformations 
(6) on the conserved (anti-)BRST charges (11). Fifth, there is a ghost-scale  symmetry (and corresponding conserved charge) in our theory 
and the (anti-)BRST and ghost charges obey the {\it standard} BRST algebra
(see, e.g. [19] for details) establishing  that the (anti-)BRST charges have the ghost numbers $(-\,1)\,1$, respectively. Sixth, it is the existence of the CF-type restriction(s) that characterizes [22, 23] a {\it quantum} version of a {\it classical}
 gauge theory discussed and described within the framework of BRST formalism. In our present {\it trivial} Abelian gauge theory, we have the {\it trivial} CF-type restriction as: $b + \bar b = 0$ where, in general, we have:
$s_{ab} c = i\,\bar b$ and $s_{b} \bar c = i\,b$.   Finally, we have seen that the BRST quantization of  our model is 
straightforward  when we take into account {\it only} the infinitesimal version of the {\it classical}
gauge transformations for our whole discussion.

\section {Nilpotent (Anti-)BRST Symmetries for the Target Space Variables and CF-Type Restriction: MBTSA}

In the previous section, we have discussed the nilpotent (anti-)BRST symmetries, conserved 
(anti-)BRST charges and BRST quantization of our model by
exploiting the beauty of the infinitesimal  {\it classical} gauge symmetry transformations:
 $\delta_g\, x_\mu = \xi\,p_\mu,\, \delta_g p_\mu = 0,\,\delta_g e = \dot\xi$.
The purpose of our present section is to exploit the infinitesimal reparameterization symmetries: 
$\delta _r\, x_\mu = \epsilon \,\dot x_\mu, \, \delta _r\, p_\mu = \epsilon  \,\dot p_\mu,\,\delta _r\,e = \frac {d}{d\tau}\, (\epsilon  \,e)$
for the discussion of the corresponding (anti-)BRST symmetries and (anti-)BRST charges in the context of
  the BRST quantization of our 1D model of a reparameterization invariant {\it free} scalar
relativistic particle. It is self-evident that the (anti-)BRST symmetry transformations for the target space variables $x_\mu (\tau)$ 
and $p_\mu (\tau)$ and einbein variable are:
$s_{ab}\,x_\mu = \bar C\, \dot x_\mu, \; s_{ab}\,p_\mu = \bar C\, \dot p_\mu, 
\; s_{ab}\,e = \frac {d}{d\tau}\,(\bar C\,e),\; s_b\,x_\mu = C\, \dot x_\mu, \; s_b\,p_\mu = C\, \dot p_\mu,
 \;s_b\,e = \frac {d}{d\tau}\,(C\,e)$ where $(\bar C)C$ are the {\it fermionic} ($C^2 = \bar C^2 = 0, \, C\, \bar C + \bar C\, C = 0$) (anti-)ghost variables corresponding to the infinitesimal parameter $\epsilon (\tau)$ present in $\tau \longrightarrow \tau - \epsilon (\tau)$. In this section, we derive the off-shell nilpotent  
(anti-)BRST symmetries for the target space variables $x_\mu (\tau)$
and $p_\mu (\tau)$ by using the {\it modified} BT-supervariable approach (MBTSA) to BRST formalism  
 [20, 21] where the super diffeomorphism transformations [cf. Eq. (13) below] 
{\it and} the {\it full } super expansions of the supervariables  along {\it all} the possible Grassmannian 
directions of the (1, 2)-dimensional supermanifold  are taken into account.

To derive the (anti-)BRST symmetries for the target space phase variables 
[i.e. $x_\mu (\tau)$ and $p_\mu (\tau)$], first of all, we {\it generalize}
the reparameterization (i.e. diffeomorphism) symmetry transformation parameter $\tau$   
[i.e. $\tau  \longrightarrow \tau^\prime  = f(\tau) \equiv \tau - \epsilon (\tau)$] from the ordinary 1D spacetime manifold 
{\it onto} our suitably  chosen  (1, 2)-dimensional supermanifold as
\begin{eqnarray}
f(\tau) \longrightarrow \tilde f(\tau, \theta, \bar\theta)  
=  \tau - \theta\;\bar C (\tau) - \bar\theta\; C(\tau) +  \theta\,\bar\theta\,h (\tau),
\end{eqnarray} 
where the supermanifold is parameterized by $(\tau, \theta, \bar\theta)$ and  we have replaced the infinitesimal 
parameter $\epsilon (\tau)$ by the {\it fermionic} (anti-)ghost variables $(\bar C)C$ 
and they have been incorporated into (13) as the coefficients of 
$(\theta)\bar\theta$ due to the fact that the Grassmannian translational generators $(\partial_\theta)\partial_{\bar\theta} $
 [along the $(\theta, \bar\theta)$-directions] have been shown [9, 10]
to be intimately connected with the nilpotent (anti-)BRST symmetry transformations
 $s_{(a)b}$. In other words, we have already incorporated  the
(anti-)BRST symmetry transformations $s_{ab} \tau  = -\,\bar C$ and $s_b\,\tau   = -\,C$ into the expansion (13). We have to compute the {\it exact} expression  for the {\it secondary} variable $h (\tau)$ from {\it other} consistency  considerations.

According  to the basic tenets of the {\it modified} BT-supervariable approach to BRST formalism, all the {\it ordinary} variables of
the theory have to be generalized   {\it onto} the suitably chosen (1, 2)-dimensional supermanifold {\it as} the supervariables where the {\it generalization} in (13) has to be incorporated as {\it one} of the arguments of the supervariables. After that, we have to take into account the {\it full} super expansions along
{\it all} the possible Grassmannian directions of the (1, 2)-dimensional supermanifold. 
Thus, we have the following explicit generalizations for the target space variables   
\begin{eqnarray}
x_\mu(\tau) &\longrightarrow & \tilde{X}_\mu(\tilde{f}(\tau, \theta, \bar{\theta}), \theta, \bar{\theta})\nonumber\\
 &=& X_\mu(\tilde{f}(\tau, \theta, \bar{\theta})) + \theta\,\bar{R}_\mu(\tilde{f}(\tau, \theta, \bar{\theta})) +  \bar{\theta}\,{R}_\mu(\tilde{f}(\tau, \theta, \bar{\theta})) + \theta\,\bar{\theta}\,S_\mu(\tilde{f}(\tau, \theta, \bar{\theta})),\nonumber\\
p_\mu(\tau) & \longrightarrow & \tilde{P}_\mu(\tilde{f}(\tau, \theta, \bar{\theta}), \theta, \bar{\theta}) \nonumber\\
&=& P_\mu(\tilde{f}(\tau, \theta, \bar{\theta})) + \theta\,\bar{T}_\mu(\tilde{f}(\tau, \theta, \bar{\theta})) +  \bar{\theta}\,{T}_\mu(\tilde{f}(\tau, \theta, \bar{\theta})) + \theta\,\bar{\theta}\,U_\mu(\tilde{f}(\tau, \theta, \bar{\theta})),
\end{eqnarray}
where {\it all} the  secondary supervariables on the r.h.s. (i.e. $R_\mu, \bar R_\mu, S_\mu, T_\mu, \bar T_\mu, U_\mu$),  are function of the {\it super} diffeomorphism transformation (13). Thus, we have to take the appropriate Taylor expansion of all the  above supervariables as: 
\begin{eqnarray}
&&X_\mu(\tau - \theta\,\bar{C} - \bar{\theta}\,C + \theta\,\bar{\theta}\,h)  =  x_\mu(\tau) - \theta\,\bar{C}\,\dot{x}_\mu 
- \bar{\theta}\,C\,\dot{x}_\mu +  \theta\,\bar{\theta}\,(h\,\dot{x}_\mu - \bar{C}\,C\,\ddot{x}_\mu), \nonumber \\
&&\theta\,\bar{R}_\mu(\tau - \theta\,\bar{C} - \bar{\theta}\,C + \theta\,\bar{\theta}\,h) = \theta\,\bar{R}_\mu(\tau) - \theta\,\bar{\theta}\,C\,\dot{\bar{R}}_\mu (\tau), \nonumber \\
&&\bar{\theta}\,{R}_\mu(\tau - \theta\,\bar{C} - \bar{\theta}\,C + \theta\,\bar{\theta}\,h) = \bar{\theta}\,{R}_\mu(\tau) + \theta\,\bar{\theta}\,\bar{C}\,\dot{{R}}_\mu (\tau), \nonumber \\
&&\theta\,\bar{\theta}\,S_\mu(\tau - \theta\,\bar{C} - \bar{\theta}\,C + \theta\,\bar{\theta}\,h) = \theta\,\bar{\theta}\,S_\mu(\tau). 
\end{eqnarray}
In the above, we have taken into account the usual key properties of the Grassmannian variables $(\theta, \bar{\theta})$ 
as: $\theta^2 = \bar{\theta}^2 = 0,          \theta\,\bar{\theta} + \bar{\theta}\,\theta = 0$. In exactly similar fashion, 
we have to expand the secondary supervariables in the expansion for
$\tilde P_\mu (\tilde f(\tau, \theta,\bar\theta), \theta, \bar\theta)$. In other words, we have the following:
\begin{eqnarray}
&&P_\mu(\tau - \theta\,\bar{C} - \bar{\theta}\,C + \theta\,\bar{\theta}\,h) = p_\mu(\tau) - \theta\,\bar{C}\,\dot{p}_\mu 
- \bar{\theta}\,C\,\dot{p}_\mu  + \theta\,\bar{\theta}\,(h\,\dot{p}_\mu - \bar{C}\,C\,\ddot{p}_\mu), \nonumber \\
&&\theta\,\bar{T}_\mu(\tau - \theta\,\bar{C} - \bar{\theta}\,C + \theta\,\bar{\theta}\,h) = \theta\,\bar{T}_\mu(\tau) - \theta\,\bar{\theta}\,C\,\dot{\bar{T}}_\mu (\tau), \nonumber \\
&&\bar{\theta}\,{T}_\mu(\tau - \theta\,\bar{C} - \bar{\theta}\,C + \theta\,\bar{\theta}\,h) = \bar{\theta}\,{T}_\mu(\tau) + \theta\,\bar{\theta}\,\bar{C}\,\dot{{T}}_\mu (\tau), \nonumber \\
&&\theta\,\bar{\theta}\,U_\mu(\tau - \theta\,\bar{C} - \bar{\theta}\,C + \theta\,\bar{\theta}\,h) = \theta\,\bar{\theta}\,U_\mu(\tau).
\end{eqnarray}
Ultimately, the {\it secondary} supervariables on the r.h.s. of (14) have to
 be replaced by the {\it ordinary } secondary variables because they are Lorentz 
{\it scalars} w.r.t. the 1D  spacetime manifold (i.e. 1D trajectory of the motion
 of the scalar relativistic particle which is embedded in 
a D-dimensional {\it target} flat Minkowskian spacetime manifold). As a consequence,
 the {\it final} expressions for the super expansions (14) are: 
\begin{eqnarray}
\tilde{X}_\mu(\tilde{f}(\tau, \theta, \bar{\theta}), \theta, \bar{\theta})  &=& x_\mu(\tau) + \theta\,\bar{R}_\mu (\tau) 
+ \bar{\theta}\,{R}_\mu (\tau)+  \theta\,\bar{\theta}\,S_\mu (\tau),\nonumber\\
&\equiv & x_\mu(\tau) + \theta\,(s_{ab}\,x_\mu (\tau))  +  \bar{\theta}\,( s_b\,x_\mu (\tau)) +  \theta\,\bar{\theta}\,(s_b\,s_{ab}\,x_\mu (\tau)),\nonumber\\
\tilde{P}_\mu(\tilde{f}(\tau, \theta, \bar{\theta}), \theta, \bar{\theta})  &=& p_\mu(\tau) + \theta\,\bar{T}_\mu (\tau) 
+ \bar{\theta}\,{T}_\mu (\tau) +  \theta\,\bar{\theta}\,U_\mu (\tau),\nonumber\\
&\equiv & p_\mu(\tau) + \theta\,(s_{ab}\,p_\mu (\tau)) + \bar{\theta}\,( s_b\,p_\mu (\tau)) +     \theta\,\bar{\theta}\,(s_b\,s_{ab}\,p_\mu (\tau)).
\end{eqnarray} 
It is clear that we have to compute explicitly the {\it exact} values of the 
secondary variables \big [$R_\mu (\tau), \bar R_\mu (\tau), S_\mu (\tau),
T_\mu (\tau), \bar T_\mu (\tau), U_\mu (\tau)$\big] for the derivation of 
the nilpotent (anti-)BRST symmetry transformations $s_{(a)b}$ for $x_\mu (\tau)$ and $p_\mu (\tau)$.
As a side remark, we would like to lay emphasis on the fact that, for the existence of the {\it proper} (anti-)BRST symmetries, we should have 
$s_b\,s_{ab}\,x_\mu (\tau)  = - \,s_{ab}\,s_{b}\,x_\mu (\tau)$ and $s_b\,s_{ab}\,p_\mu (\tau)  = - \,s_{ab}\,s_{b}\,p_\mu (\tau)$
which lead to the absolute anticommutativity (i.e. $s_b\,s_{ab} + s_{ab}\,s_b = 0$) of the (anti-)BRST symmetry
 transformations  [$s_{(a)b}$].

At this stage, we exploit the theoretical potential and  power of the ``horizontality condition" (HC) for the reparameterization invariant theory and demand the following on physical ground\footnote{A (fermionic and/or bosonic) Lorentz  scalar 
should remain the {\it same} Lorentz scalar under any kind of spacetime and/or internal, (non-)supersymmetric, etc.,
transformations.}
\begin{eqnarray}
&& \tilde{X}_\mu(\tilde{f}(\tau, \theta, \bar{\theta}), \theta, \bar{\theta})   \equiv {X}_\mu(\tau, \theta, \bar{\theta})= x_\mu (\tau), \nonumber\\
&& \tilde{P}_\mu(\tilde{f}(\tau, \theta, \bar{\theta}),\theta, \bar{\theta})   \equiv {P}_\mu(\tau, \theta, \bar{\theta})= p_\mu (\tau),
\end{eqnarray} 
due to the fact that $x_\mu (\tau) $ and  $p_\mu (\tau) $ are {\it scalars} w.r.t. the 1D diffeomorphism transformation.
For the above equality to be true, we have to collect {\it all} the expansions in (15) and (16) in a systematic and precise  manner as illustrated below, namely: 
\begin{eqnarray}
\tilde{X}_\mu(\tilde{f}(\tau, \theta, \bar{\theta}), \theta, \bar{\theta})   &=&  x_\mu (\tau) + \theta\, (\bar R_\mu - \bar C\,\dot x_\mu) + \bar\theta\, (R_\mu - C\,\dot x_\mu)\nonumber\\
& + &   \theta\,\bar\theta\,\big [ S_\mu + \bar C\,\dot{R}_\mu -  C\,\dot{\bar R}_\mu + h\; \dot x_\mu  - \bar C\,C\,\ddot x_\mu \big],\nonumber\\
 \tilde{P}_\mu(\tilde{f}(\tau, \theta, \bar{\theta}), \theta, \bar{\theta})   & = & p_\mu (\tau) + \theta\, (\bar T_\mu - \bar C\,\dot p_\mu)  + \bar\theta\, (T_\mu - C\,\dot p_\mu)\nonumber\\
& + &   \theta\,\bar\theta\,\big [ U_\mu + \bar C\,\dot{T}_\mu -  C\,\dot{\bar T}_\mu + h \; \dot p_\mu  - \bar C\,C\,\ddot p_\mu \big].
\end{eqnarray}
Now, we utilize the theoretical potential and power of the HC. Mathematically, this requires that:
$\tilde{X}_\mu(\tilde{f}(\tau, \theta, \bar{\theta}), \theta, \bar{\theta})   
= x_\mu (\tau),  \,\tilde{P}_\mu(\tilde{f}(\tau, \theta, \bar{\theta}), \theta, \bar{\theta})   = p_\mu (\tau)$. This 
leads to the determination of the {\it secondary} variables as
\begin{eqnarray}
&&\bar{R}_\mu = \bar{C}\,\dot{x}_\mu,\qquad R_\mu = C\,\dot{x}_\mu, \qquad  S_\mu = C\,\dot{\bar{R}}_\mu - \bar{C}\,\dot{R}_\mu + \bar{C}\,C \ddot{x}_\mu - \; h\, \dot x_\mu, \nonumber \\
&&\bar{T}_\mu = \bar{C}\,\dot{p}_\mu, \qquad T_\mu = C\,\dot{p}_\mu,\qquad U_\mu = C\,\dot{\bar{T}}_\mu - \bar{C}\,\dot{T}_\mu + \bar{C}\,C \,\ddot{p}_\mu - \, h \, \dot p_\mu.
\end{eqnarray}
Plugging in the values of $R_\mu, \bar{R}_\mu, T_\mu$ and $\bar{T}_\mu$ in the above, we obtain the following expressions for $S_\mu(\tau)$ and $U_\mu(\tau)$, namely:
\begin{eqnarray} 
&&S_\mu(\tau) = -[(\dot{\bar{C}}\,C + \bar{C}\,\dot{C} + h)\,\dot{x}_\mu + \bar{C}\,C\,\ddot{x}_\mu], \nonumber \\
&&U_\mu(\tau) = -[(\dot{\bar{C}}\,C + \bar{C}\,\dot{C} + h)\,\dot{p}_\mu + \bar{C}\,C\,\ddot{p}_\mu].
\end{eqnarray}
As argued earlier, the above expressions are {\it also} equal to $s_b\,s_{ab}\,x_\mu \equiv -s_{ab}\,s_b\,x_\mu$ and 
$s_b\,\,s_{ab}\,\,p_\mu = -s_{ab}\,s_b\,\,p_\mu$, respectively,  where we have already derived $s_b\,\,x_\mu = 
C\,\dot{x}_\mu, s_{ab}\,\,x_\mu = \bar{C}\,\dot{x}_\mu, s_b\,\,p_\mu = C\,\dot{p}_\mu$ and $s_{ab}\,\,p_\mu = 
\bar{C}\,\dot{p}_\mu$ because of the comparison with $(17)$. In other words, as is evident from $(20)$, the 
expressions for $R_\mu, \bar{R}_\mu, T_\mu$ and $\bar{T}_\mu$ imply that we have {\it already} obtained  the nilpotent (anti-)BRST symmetries $s_{(a)b}$ for the target space variables $x_\mu(\tau)$ and $p_\mu(\tau)$.

The nilpotency $[s_{(a)b}^2 = 0]$ properties of $s_{(a)b}$ lead to the derivation 
of the (anti-) BRST symmetry transformations on the (anti-)ghost variables as:
\begin{eqnarray}
s_b\,C = C\,\dot{C}, \qquad\qquad s_{ab}\,\bar{C} = \bar{C}\,\dot{\bar{C}}.
\end{eqnarray}  
We assume that $s_{ab}\,C = \bar{B}$ and $s_b\,\bar{C} = B$ where $B$ and $\bar{B}$ are the Nakanishi-Lautrup type auxiliary variables of the theory. These transformations (i.e. $s_b\,\bar{C} = B, s_{ab}\,\,C = \bar{B}$) are the {\it standard} assumptions in the realm of BRST formalism. As a consequence of these
off-shell nilpotent [i.e. $ s_{(a)b} ^2 = 0]$  transformations $(s_b\,B = 0, s_{ab}\,\bar B= 0$), we note the following:
\begin{eqnarray}
s_b\,s_{ab}\,x_\mu & = & (B - \bar{C}\,\dot{C})\,\dot{x}_\mu - \bar{C}\,C\,\ddot{x}_\mu \equiv S_\mu (\tau), \nonumber \\
- s_{ab}\,s_{b}\,x_\mu & = &  (- \bar{B} - \dot{\bar{C}}\,{C})\,\dot{x}_\mu - \bar{C}\,C\,\ddot{x}_\mu \equiv S_\mu (\tau), \nonumber \\
s_b\,s_{ab}\,p_\mu & = & (B - \bar{C}\,\dot{C})\,\dot{p}_\mu - \bar{C}\,C\,\ddot{p}_\mu \equiv U_\mu (\tau), \nonumber \\
- s_{ab}\,s_{b}\,p_\mu & = & (- \bar{B} - \dot{\bar{C}}\,{C})\,\dot{p}_\mu - \bar{C}\,C\,\ddot{p}_\mu \equiv U_\mu (\tau).
\end{eqnarray}
The comparison of the above  {\it with} the expressions (21) (derived from the MBTSA) leads to the derivation  of the secondary variable $h(\tau)$ as
\begin{eqnarray}  
h(\tau) & = & \bar{B} - \bar{C}\,\dot{C}  \equiv  - B - \dot{\bar{C}}\,C \quad  \Longrightarrow \quad   B + \bar{B} + (\dot{\bar{C}}\,C - \bar{C}\,\dot{C}) = 0.
\end{eqnarray}
Thus, we have derived the celebrated Curci-Ferrari (CF)-type restriction
 \big[i.e. $B + \bar B + (\dot{\bar C}\,C - \bar C\,\dot C) = 0 $\big]
from the application of MBTSA to BRST formalism where it is the determination
 of the secondary variable $h (\tau)$ [cf. Eq. (13)], in terms of the {\it basic}
and {\it auxiliary} variables, from the requirement (i.e. $s_b\,s_{ab}\,x_\mu = -\,s_{ab} \,s_b\,x_\mu$ or $s_b\,s_{ab}\,p_\mu 
= -\,s_{ab} \,s_b\,p_\mu$) of the absolute anticommutativity (i.e. $s_b\,s_{ab} + s_{ab} \,s_b = 0$) 
property of $ s_{(a)b}$ that has played a crucial role.

We end this section with the following remarks. First, we note that our choice of $s_b\, \bar C = B$ and  $s_{ab}\,C = \bar B$ 
implies that we have the following generalizations for the (anti-)ghost variables 
$(\bar C) C$ from the 1D {\it ordinary} spacetime manifold to 
the (1, 1)-dimensional {\it (anti-)chiral} super submanifolds of the (1, 2)-dimensional supermanifold, as  
\begin{eqnarray}  
C (\tau) \longrightarrow F^{(c)} (\tau, \theta) &=& C(\tau) + \theta\, [\bar B (\tau)]\equiv   C (\tau) + \theta\, (s_{ab}C),\nonumber\\
~~~~\bar C (\tau) \longrightarrow \bar F^{(ac)} (\tau, \bar\theta) & = & \bar C(\tau) + \bar\theta\, [B (\tau)] \equiv   \bar C (\tau)  + \bar\theta\, (s_{b}\bar C),
\end{eqnarray}
where the superscripts $(c)$ and $(ac)$ denote the {\it chiral} and {\it anti-chiral} 
super expansions. This observation, in a {\it subtle}  manner, explains that the (anti-)chiral
 supervariable approach (ACSA) to BRST formalism [15-18] would be useful to us in our further discussions.
Second, it can be checked that the absolute 
anticommutativity ($s_b\, s_{ab} + s_{ab}\, s_b = 0$) properties, for the phase space target variables 
[i.e. $x_\mu (\tau), \,p_\mu (\tau)$] w.r.t. the off-shell nilpotent 
(anti-)BRST symmetry   transformations, namely;
\begin{eqnarray}
&&\{s_b, s_{ab}\}\, x_\mu = [B + \bar B + (\dot{\bar C}\,C - \bar C\,\dot C)]\;\dot x_\mu = 0,\nonumber\\
&&\{s_b, s_{ab}\}\, p_\mu = [B + \bar B + (\dot{\bar C}\,C - \bar C\,\dot C)]\;\dot p_\mu = 0,
\end{eqnarray}
 are valid if and only if we apply the power  and  potential  of the CF-type restriction (24) from {\it outside}. 
Finally, we note that the requirement of the absolute anticommutativity of the (anti-)BRST symmetry transformations $s_{(a)b}$ on the
 (anti-)ghost variables, namely;
\begin{eqnarray}
&&\{s_b, s_{ab}\}\,C = 0 \;\Longrightarrow \;s_b \bar B = \dot{\bar B}\;C - \bar B\,\dot C,\nonumber\\
&&\{s_b, s_{ab}\}\,\bar C = 0\;\Longrightarrow \; s_{ab}  B = \dot{B}\;\bar C -  B\,\dot{\bar C},
\end{eqnarray} 
leads to the derivation of $s_b \bar B = \dot{\bar B}\,C - \bar B\,\dot C$ and $s_{ab}  B = \dot{B}\,\bar C -  B\,\dot{\bar C}$ 
which are found to be off-shell nilpotent ($[s_{(a)b}]^2 = 0$) and absolutely anticommuting in nature 
(i.e. $\{s_b, s_{ab}\}\,B = 0,\; \{s_b, s_{ab}\}\,\bar B = 0$)
{\it without} any use of the CF-type restriction.

\section{(Anti-)BRST Symmetry  Transformations for other Variables of the Theory: ACSA}

As has been pointed out earlier, we have already utilized the {\it (anti-)chiral} supervariable approach (ACSA) to determine the 
(anti-)BRST symmetry transformations: $s_{ab}\,C = \bar B$ and $s_{b}\,\,{\bar C} =  B$ [cf. Eq. (25)] which are {\it primarily} assumed in the BRST approach.
In this section, we apply the ACSA to BRST formalism to derive the {\it rest} of the off-shell  nilpotent (anti-)BRST symmetry 
transformations ({\it besides} our derivations
in the previous section which are: $s_b\,x_\mu = C\, \dot x_\mu,\; s_{ab}\,x_\mu = \bar C\, \dot x_\mu, \; s_b\,p_\mu = C\, \dot p_\mu,\;
s_{ab}\,p_\mu = \bar C\, \dot p_\mu,\; s_b\,\bar C  = B,\;s_{ab}\,C = \bar B$).
Towards this objective  in mind, we generalize  the 1D {\it ordinary}
variables [$e(\tau),\,C(\tau),\,\bar B(\tau),\,B(\tau)$] onto a (1, 1)-dimensional {\it anti-chiral} 
super-submanifold of the general (1, 2)-dimensional supermanifold as
\begin{eqnarray} 
&&e(\tau)\; \longrightarrow E(\tau, \bar\theta) = e(\tau) + \bar\theta\,f_1(\tau), \nonumber\\
&&\bar{B}(\tau)\; \longrightarrow \tilde{\bar{B}}(\tau, \bar\theta) = \bar B(\tau) + \bar\theta\,f_3(\tau),\nonumber\\
&&B(\tau) \;\longrightarrow \tilde{B}(\tau, \bar\theta) = B(\tau) + \bar\theta\,f_2(\tau), \nonumber\\
&& C(\tau) \longrightarrow F(\tau, \bar\theta) = C(\tau) + \bar\theta\,b_1(\tau), 
\end{eqnarray}
where the {\it secondary} variables $(f_1, f_2, f_3)$ are {\it fermionic} and $b_1$ is {\it bosonic} 
in nature because of the fermionic $(\bar{\theta}^2 = 0)$ nature of $\bar{\theta}$. It is evident that,
 in the limit $\bar{\theta} = 0$, we get back our {\it ordinary} variables $[e(\tau), C(\tau), B(\tau), 
 \bar{B}(\tau)]$ from the above super expansions. Furthermore, it should be noted that our (1, 1)-dimensional
{\it anti-chiral} super-submanifold is parameterized by $(\tau, \bar\theta)$
where the evolution parameter  $\tau$ is {\it bosonic} and $\bar\theta$ is {\it fermionic}.

To determine the {\it secondary} variables, in terms of the {\it basic} and {\it auxiliary} variables of 
the theory, we have to exploit {\it one} of the basic tenets of the ACSA which states that the {\it quantum}
 gauge (i.e. BRST) invariant quantities should be independent of the Grassmannian variable $\bar{\theta}$. 
In this context, we note:
\begin{eqnarray}
&& s_b\,(C\,\dot{x}_\mu) = 0, \quad s_b\,(e\,\dot{C} + \dot{e}\,C) = 0, \quad  s_b\,(\dot{\bar{B}}\,C - \bar{B}\,\dot{C}) = 0, \quad s_b\,B = 0.
\end{eqnarray}      
The above interesting  BRST-invariant quantities, generalized onto a $(1, 1)$-dimensional 
 {\it anti-chiral} super-submanifold, should be independent of $\bar{\theta}$. In other words, we have the validity of the  following
equalities, namely; 
\begin{eqnarray}
&&F(\tau, \bar{\theta})\,\dot{X}_\mu^{(ha)}(\tau, \bar{\theta}) = C(\tau)\,\dot{x}_\mu(\tau), \quad\tilde{B}(\tau, \bar{\theta}) = B(\tau), \nonumber\\
&& E(\tau, {\bar{\theta}})\,\dot{F}(\tau, \bar{\theta})  + \dot{E}(\tau, \bar{\theta})\,F(\tau, \bar{\theta}) 
= e(\tau)\,\dot{C}(\tau) + \dot{e}(\tau)\,C(\tau),\nonumber\\
&& \dot{\tilde{\bar{B}}}(\tau, \bar{\theta}) \, F(\tau, \bar{\theta}) - \tilde{\bar{B}}(\tau, \bar{\theta})\,\dot F(\tau, \bar{\theta})
 = \dot{\bar{B}}(\tau)\,C(\tau) - \bar{B}(\tau)\,\dot{C}(\tau), 
\end{eqnarray}
where $X_\mu^{(ha)}(\tau, \bar{\theta})$ is the {\it anti-chiral} limit of the {\it full} super expansion that has been obtained in the {\it previous} section, namely;
\begin{eqnarray}
X_\mu^{(h)}(\tau, \theta, \bar{\theta}) & = &  x_\mu(\tau) + {\theta}\,(\bar C\,\dot{x}_\mu) + \bar\theta \,(C\, \dot x_\mu) 
 +  \theta\,\bar\theta\,\big[-\{(\bar B + \dot {\bar C}\,C)\,\dot x_\mu + \bar C\,C\, \ddot x_\mu\}\big]\nonumber\\
&\equiv & x_\mu(\tau) + {\theta}\,(\bar C\,\dot{x}_\mu ) + \bar\theta \,(C\, \dot x_\mu )
+   \theta\,\bar\theta\;\big[(B - \bar C\,\dot C)\,\dot x_\mu  - \bar C\,C\, \ddot x_\mu \big].
\end{eqnarray}
In the above,  the superscript $(h)$ denotes that the supervariable $X_\mu ^{(h)} (\tau, \theta, \bar\theta)$ has been
 obtained after the application of the HC. In other words, we have the following {\it anti-chiral} limiting case, namely,  
\begin{eqnarray}
X_\mu^{(ha)}(\tau, \bar{\theta}) &=& x_\mu(\tau)  + \bar\theta \,[C (\tau)\, \dot x_\mu (\tau)],
\end{eqnarray}
where the superscript $(ha)$ denotes the {\it anti-chiral} limit of the super expansion (31) that has been obtained after the application of the HC
in the previous section. The substitutions, from $(31)$ and $(28)$ into the {\it first} entry of Eq. (30), 
leads to $b_1 (\tau) = C (\tau)\,\dot C (\tau)$. The BRST invariance of the Nakanishi-Lautrup auxiliary variable (i.e. $s_b\,B  = 0$)
implies that $f_2 (\tau) = 0$. Thus, we have the following super expansions:
\begin{eqnarray} 
&&F ^{(b)}(\tau, \bar\theta)  = C(\tau) + \bar\theta\,(C\,\dot C) \;\equiv \; C(\tau) + \bar\theta\,(s_{b}\,C (\tau)), \nonumber\\
&&\tilde{B} ^{(b)}(\tau, \bar\theta) = B(\tau) + \bar\theta\,(0)\; \equiv \; B(\tau) + \bar\theta\,(s_{b}\,B (\tau)).
\end{eqnarray}
A close look at the above equation demonstrates that we have already obtained the BRST symmetry transformations:
 $s_b\,C = C\,\dot C$ and $s_b\,B = 0$ as the coefficients of $\bar\theta$ in the 
expansions (33) where the superscript $(b)$ denotes the supervariables that have been obtained after the applications of the BRST-invariant 
(i.e. {\it quantum } gauge invariant) restrictions (29). It should be noted that $s_b\,C = C\,\dot C$ can {\it also} be derived from the 
{\it restriction} corresponding to the invariance $s_b\,(C\,\dot p_\mu) = 0$ on the (1, 1)-dimensional {\it anti-chiral} super-submanifold.
However, for the sake of brevity, we have {\it not} discussed {\it it} here. In the rest of the restrictions in (30), we use the {\it final}
expressions from (33) to obtain the {\it exact} expressions for the secondary {\it variables} as:
\begin{eqnarray} 
&&f_1 (\tau) = e (\tau)\,\dot C(\tau) + \dot e (\tau)\,C(\tau), \nonumber\\
&& f_3 (\tau) = \dot {\bar B} (\tau)\, C(\tau) - \bar B (\tau)\,\dot C (\tau).
\end{eqnarray}
As a consequence, we have the following super expansions for {\it some} of the supervariables [cf. Eq. (28)],  namely; 
\begin{eqnarray} 
&&E ^{(b)}(\tau, \bar\theta)  = e(\tau) + \bar\theta\,(\dot e\,C + e\,\dot C) \;\equiv \; e (\tau) + \bar\theta\,(s_{b}\,e (\tau)), \nonumber\\
&&\tilde{\bar B} ^{(b)}(\tau, \bar\theta) = \bar B(\tau) + \bar\theta\,(\dot {\bar B}\, C - \bar B\,\dot C)\; \equiv \; \bar B(\tau) + \;\bar\theta\,(s_{b}\,\bar B (\tau)),
\end{eqnarray}
where the superscript $(b)$ stands for the  expansions that have been obtained after the applications of the BRST
 (i.e. {\it quantum} gauge) invariance listed in $(29)$.
It is straightforward to note that we have already derived the BRST transformations: $s_b\,B = 0, s_b\,C = 
C\,\dot{C}, s_b\,e = e\,\dot{C} + \dot{e}\,C$ and $s_b\,\bar{B} = \dot{\bar{B}}\,C - \bar{B}\,\dot{C}$ as 
the coefficients of $\bar{\theta}$-Grassmannian variable in the super expansions of equations (33) and (35).
 In other words, we observe that {\it all} the BRST symmetry transformations $(s_b)$ for {\it all} the variables of our
 theory have been obtained in equations (25), (33) and (35) {\it besides} the target space variables that have 
been obtained earlier by exploiting the theoretical strength of MBTSA (cf. Sec. 3).

To derive the anti-BRST symmetry transformations $(s_{ab})$ for the variables $(B, e, \bar{C}, \bar{B})$,
 we note that the following quantities 
(that are present in the round brackets) are anti-BRST invariant, namely;
\begin{eqnarray}
&&s_{ab}\,\bar{B} = 0, \;\; s_{ab}\,(\dot{B}\,\bar{C} - B\,\dot{\bar{C}}) = 0, \quad s_{ab}\,(e\,\dot{\bar{C}}  + \dot{e}\,\bar{C}) = 0, \;\; s_{ab}\,(\bar{C}\,\dot{x}_\mu) = 0.
\end{eqnarray} 
According to the basic tenets of ACSA, the above quantities {\it must} be independent of the Grassmannian
 variable $(\theta)$ when they are generalized onto a $(1,1)$-dimensional {\it chiral} super-submanifold
 of the $(1,2)$-dimensional supermanifold on which our theory is generalized. Towards this aim   in mind,
 we generalize the 1D variables $(e, B, \bar{B}, \bar{C})$ onto the chosen $(1,1)$-dimensional {\it chiral}
 super-submanifold as the following super expansions, namely;
\begin{eqnarray}
&&e(\tau) \longrightarrow E(\tau, \theta) = e(\tau) + \theta\,\bar{f}_1(\tau), \nonumber\\
&& B (\tau) \longrightarrow \tilde{B}(\tau, \theta) = B(\tau) + \theta\,\bar{f}_2(\tau), \nonumber \\
&&\bar{C}(\tau) \longrightarrow \bar{F}(\tau, \theta) = \bar{C}(\tau) + \theta\,\bar{b}_1(\tau), \nonumber\\
&&\bar{B} (\tau) \longrightarrow \tilde{\bar{B}}(\tau, \theta) = \bar{B}(\tau ) + \theta\,\bar{f}_3(\tau), 
\end{eqnarray}
where $(\bar{f}_1, \bar{f}_2, \bar{f}_3)$ are {\it fermionic} and $\bar{b}_1(\tau)$ is the 
{\it bosonic} secondary variables because of the fermionic $(\theta^2 = 0)$ nature of the
 Grassmannian variable  $\theta$ which characterizes the $(1, 1)$-dimensional {\it chiral} 
super-submanifold {\it besides} the evolution bosonic parameter $\tau$
of our 1D diffeomorphism invariant system.

As a first-step, let us compute  the secondary variable $\bar b_1(\tau)$ in terms of the 
{\it basic} variables of the theory. The anti-BRST invariance we use is: $s_{ab}\,(\bar{C}\,\dot{x}_\mu)
 = 0$. In other words, we have the validity of the following restriction on the {\it chiral} supervariables according to
 the basic tenets of ACSA, namely;
\begin{eqnarray}
\bar F (\tau, \theta)\,\dot X_\mu ^{(hc)} (\tau, \theta) = \bar C (\tau)\, \dot x_\mu (\tau),
\end{eqnarray}
where $X_\mu ^{(hc)}$  is the {\it chiral} limit of the {\it full} super expansion (31) 
that has been obtained for $X_\mu ^{(h)} (\tau, \theta, \bar\theta)$. To be precise, the {\it latter} 
has been derived in the previous section. Mathematically, the above {\it chiral} limit  implies the following:
\begin{eqnarray}
X_\mu ^{(hc)} (\tau, \theta) = x_\mu (\tau) + \theta \,(\bar C(\tau)\,\dot x _\mu (\tau)).
\end{eqnarray} 
Plugging in the expansions  from (37) and (39), we obtain the expression for $\bar b_1 (\tau) = 
\bar C (\tau) \,\dot{\bar C}(\tau)$. Thus, we have already obtained
\begin{eqnarray}
\bar F ^{(ab)}  {(\tau, \theta)}= \bar C (\tau) + \theta\,(\bar C\,\dot {\bar C}) \equiv \bar C (\tau) + \theta\, (s_{ab}\bar C (\tau)),
\end{eqnarray} 
where the coefficient of $\theta$ is nothing but the anti-BRST symmetry transformation for the $\bar C (\tau)$ variable 
as\footnote{Exactly similar kinds of exercise can be performed with the variable $p_\mu (\tau)$ and, from the restriction $s_{ab}\,(\bar C\,\dot p_\mu) = 0$,
we can obtain the anti-BRST symmetry transformation of the anti-ghost variable as: $s_{ab}\,\bar C = \bar C\,\dot{\bar C}$. 
For the sake of brevity, however, we have {\it not} discussed it explicitly here.}: 
$s_{ab} \bar C = \bar C\,\dot {\bar C}$ and the superscript $(ab)$ denotes the supervariable that has been obtained after the application of the  specific anti-BRST invariant restriction in (36).
Against the backdrop of the above derivation, we can derive the {\it other} anti-BRST symmetry transformations by using the anti-BRST (i.e. {\it quantum} gauge) invariant quantities $(36)$ and using the super expansions $(37)$ and $(40)$. In other words, we have the following restrictions
\begin{eqnarray}   
&&\tilde{\bar{B}}(\tau, \theta) = \bar{B}(\tau), \quad 
\dot {\tilde B} (\tau, \theta)\, \bar F ^{(ab)} (\tau, \theta)- {\tilde B (\tau, \theta)}\,\dot {\bar F} ^{(ab)} (\tau, \theta) = \dot B (\tau)\,\bar C (\tau) - \,B (\tau)\, \dot {\bar C} (\tau),\nonumber\\
&& E (\tau, \theta)\, \dot {\bar F }^{(ab)} (\tau, \theta) + \dot E (\tau, \theta)\,\bar F ^{(ab)} (\tau, \theta) = e (\tau)\,\dot{\bar C} (\tau) + \, \dot e (\tau)\, \bar C (\tau), 
\end{eqnarray}
which lead to the precise determination  of the secondary variables as follows:
\begin{eqnarray}
&&\bar{f}_3(\tau) = 0, \quad  \bar f_1 (\tau) = e(\tau)\,\dot {\bar C} + \dot e (\tau)\, \bar C (\tau), \;\;\bar f_2 (\tau) = \dot B (\tau)\, \bar C (\tau) - B (\tau)\, \dot {\bar C} (\tau).
\end{eqnarray}
Ultimately, we have the following super expansions in their full blaze of glory
\begin{eqnarray} 
&&X_\mu^{(hc)}(\tau, \theta) = x_\mu(\tau) + \theta\,(\bar{C}\,\dot{x}_\mu) \equiv x_\mu(\tau) + \theta\,(s_{ab}\,x_\mu(\tau)), \nonumber \\
&&P_\mu^{(hc)}(\tau, \theta) = p_\mu(\tau) + \theta\,(\bar{C}\,\dot{p}_\mu) \equiv p_\mu(\tau) + \theta\,(s_{ab}\,p_\mu(\tau)), \nonumber \\
&&E^{(ab)}(\tau, \theta) = e(\tau) + \theta\,(e\,\dot{\bar{C}} + \dot{e}\,\bar{C}) \equiv e(\tau) + \theta\,(s_{ab}\,e(\tau)), \nonumber\\
&&F^{(ab)}(\tau, \theta)  = C(\tau) + \theta\,(\bar{B}) \equiv  C(\tau) + \theta\,(s_{ab}\,C (\tau)), \nonumber\\
&& \bar{F}^{(ab)}(\tau, \theta) = \bar{C}(\tau) + \theta\,(\bar{C}\,\dot{\bar{C}}) \equiv \bar{C}(\tau) + \theta\,(s_{ab}\,\bar{C} (\tau)), \nonumber\\
&&\tilde{B}^{(ab)}(\tau, \theta) = B(\tau) + \theta\,(\dot{B}\,\bar{C} - B\,\dot{\bar{C}}) \equiv B(\tau) + \theta\,(s_{ab}\,B (\tau)), \nonumber\\
&& \tilde{\bar{B}}^{(ab)}(\tau, \theta)  = \bar{B}(\tau) + \theta\,(0) \equiv  \bar{B}(\tau) + \theta\,(s_{ab}\,\bar{B} (\tau)).
\end{eqnarray}
where the superscripts $(hc)$ and $(ab)$ have been already explained earlier. We observe that the
 coefficients of $\theta$ in (43) are nothing but the anti-BRST transformations for 
{\it all} the variables $(x_\mu, p_\mu, e, B, \bar{B}, C, \bar{C})$ of our theory 
 (cf. Secs. 3 and 4).

\section{Lagrangian Formulation: Reparameterization Symmetry and Corresponding (Anti-)BRST Symmetry Transformations}

In this section, we elevate the {\it classical} reparameterization symmetry $\tau \rightarrow {\tau}' = \tau - \epsilon(\tau)$ to its {\it quantum} counterparts  within the framework of BRST formalism. In this context, the nilpotent (anti-)BRST symmetries (that have been derived in the previous section) help in finding out the gauge-fixing and Faddeev-Popov (FP) ghost terms in the following manner\footnote{It should be noted that we have taken the {\it same} combination of variables in the square bracket $(44)$ which has been taken in Sec. 2, in the context of BRST quantization, corresponding to the {\it gauge} symmetry {\it modulo} a factor of $i$. The {\it latter} has been taken for the sake of brevity.}:
\begin{eqnarray}
-s_{ab}\,s_{b}\Big[\frac{e^2}{2} - \frac{\bar{C}\,C}{2}\Big] = \bar{B}\,\big[e\,\dot{e} + 2\,{\bar{C}}\,\dot{C} + \dot{\bar{C}}\,{C}\big]
 - \frac{\bar{B}^2}{2} -  e^2\,\dot{\bar{C}}\,\dot{C} - e\,\dot{e}\,{\bar{C}}\,\dot{C} - \dot{\bar{C}}\,\bar{C}\,\dot{C}\,C,\nonumber\\
s_b\,s_{ab}\Big[\frac{e^2}{2} - \frac{\bar{C}\,C}{2}\Big] = -B\,\big[e\,\dot{e} + 2\,\dot{\bar{C}}\,C + \bar{C}\,\dot{C}\big] - \frac{B^2}{2} - e^2\dot{\bar{C}}\,\dot{C} - e\,\dot{e}\,\dot{\bar{C}}\,C - \dot{\bar{C}}\,\bar{C}\,\dot{C}\,C.
\end{eqnarray} 
As a consequence of $(44)$, we have the following (anti-)BRST invariant coupled (but equivalent) Lagrangians for our theory, namely;
\begin{eqnarray}
 L_{\bar{B}} & = & p_\mu\,\dot{x}^\mu - \frac{e}{2}\,\big(p^2 - m^2\big) + \bar{B}\,\big(e\,\dot{e} + 2\,{\bar{C}}\,\dot{C}
 + \dot{\bar{C}}\,{C}\big)\nonumber\\
 & - & \frac{\bar{B}^2}{2} - e^2\,\dot{\bar{C}}\,\dot{C} -  e\,\dot{e}\,{\bar{C}}\,\dot{C} - \dot{\bar{C}}\,\bar{C}\,\dot{C}\,C,\nonumber\\
L_{{B}} & = & p_\mu\,\dot{x}^\mu - \frac{e}{2}\,\big(p^2 - m^2\big) -B\;\big(e\,\dot{e} + 2\,\dot{\bar{C}}\,C + \bar{C}\,\dot{C}\big)\nonumber\\
& - & \frac{B^2}{2} - e^2\,\dot{\bar{C}}\dot{C} -  e\,\dot{e}\,\dot{\bar{C}}\,C - \dot{\bar{C}}\,\bar{C}\,\dot{C}\,C.
\end{eqnarray}    
We point out that the pure FP-ghost part $(i.e.\, - \dot{\bar{C}}\,\bar{C}\,\dot{C}\,C )$ of the Lagrangians (45) remains the {\it same}. Furthermore, because of the off-shell nilpotency 
[$s_{(a)b} ^2 = 0$] of the (anti-)BRST symmetries $s_{(a)b}$, it is straightforward to note that $L_{\bar B}$
would be anti-BRST invariant and $L_B$ would be BRST invariant [cf. Eq. (44)]. To corroborate the {\it latter}
statement, we note the sanctity of the following\footnote{We are sure that $L_{\bar B}$ and $L_{B}$ would be (anti-)BRST invariant because the 
first-order Lagrangian $L_{f}$ [i.e. the {\it first} two terms of (45)] transforms to a total derivative under the infinitesimal reparameterization 
[i.e. diffeomorphism transformations (5)] (cf. Sec. 2). As a consequence, under the nilpotent [$s^{2}_{(a)b} = 0$] (anti-)BRST symmetry transformations,
  $L_{f}$ would transform as: $s_{ab}\, L_{f} = \frac {d}{d\tau} (\bar C\,L_f), \;\, s_b\, L_f   =  \frac {d}{d\tau} (C\,L_f$).}  
\begin{eqnarray}
 s_{ab}\, L_{\bar B} &=& \frac {d}{d\tau}\Big [ \bar C\,L_f + e^2\,\bar B\,\dot {\bar C} + e\,\dot e\,\bar B\, \bar C 
\,\bar C\,C -   \bar B\, \dot {\bar C} -  {\bar B}^2\,\bar C\Big],\nonumber\\
 s_b\, L_B  &=&  \frac {d}{d\tau}\Big [ C\,L_f - e^2\,B\,\dot C - e\,\dot e\,B\, C  -   B\,\bar C\,\dot C\, C -   B^2\,C\Big],
\end{eqnarray}
which render the action integrals $S_1 = \int_{-\infty }^{+\infty } d\tau \,L_{\bar B}$ and $S_2 = \int_{-\infty }^{+\infty } d\tau \,L_{B}$
 of our theory (described by the coupled Lagrangian densities  $L_{\bar B}$ and $L_{B}$)  (anti-)BRST invariant, respectively, for the 
physically well-defined variables which vanish-off as $\tau \longrightarrow  \pm\,\infty$. In the above, the first-order Lagrangian  $L_f$ is {\it same} as defined 
in Sec. 2 and the {\it full} nilpotent (anti-)BRST  transformations (for our 1D theory of a {\it scalar} relativistic particle) are as follows:
\begin{eqnarray}
&&s_{ab} x_\mu = \bar C\, \dot x_\mu, \, s_{ab} p_\mu = \bar C\, \dot p_\mu, \,  s_{ab} C = \bar B,
\;  s_{ab} \bar C = \bar C\,\dot {\bar C}, \nonumber\\
&& s_{ab} e = \frac {d}{d\tau}\,(\bar C\,e),\;\; s_{ab} \bar B = 0,\quad \; s_{ab} B = \dot{B}\,\bar C - B\,\dot {\bar C},\nonumber\\
&&s_b x_\mu = C\, \dot x_\mu, \; s_b p_\mu = C\, \dot p_\mu, \; s_b C = C\,\dot C,\;\, s_b \bar C = B, \nonumber\\
 && s_b e = \frac {d}{d\tau}\,(C\,e),\quad s_b B = 0, \quad s_b \bar B = \dot{\bar B}\,C - \bar B\,\dot C. 
\end{eqnarray}
The above transformations are off-shell nilpotent [$s_{(a)b} ^2 = 0$] and absolutely anticommuting in nature. The absolute anticommutativity 
($s_b\, s_{ab} + s_{ab}\, s_b = \{s_b, s_{ab}\} = 0)$ property is {\it true} for {\it all} variables of our theory, namely;
\begin{eqnarray}
&&\{s_b, s_{ab}\}\, x_\mu = [B + \bar B + (\dot {\bar C}\, C - \bar C\, \dot C)]\;\dot x_\mu = 0,\nonumber\\
&&\{s_b, s_{ab}\}\, p_\mu = [B + \bar B + (\dot {\bar C}\, C - \bar C\, \dot C)]\;\dot p_\mu = 0,\nonumber\\
&&\{s_b, s_{ab}\}\, e = \frac {d}{d\tau}\Big[\big\{B + \bar B + (\dot {\bar C}\, C - \bar C\, \dot C)\big\}\,e\Big] = 0,\nonumber\\
&&\{s_b, s_{ab}\}\, C = 0,\quad\{s_b, s_{ab}\}\, \bar C = 0,\nonumber\\
&&\{s_b, s_{ab}\}\, B = 0, \;\quad\{s_b, s_{ab}\}\, \bar B = 0,
\end{eqnarray}
{\it provided} we impose the (anti-)BRST invariant CF-type restriction: $B+ \bar B + \dot {\bar C}\, C - \bar C\, \dot C = 0$ from {\it outside} 
on our theory which is, obviously, a {\it physical} requirement because it remains invariant under the (anti-)BRST symmetry 
transformations (47). In other words, it can be checked that $s_{(a)b}\,[B+ \bar B + (\dot {\bar C}\, C - \bar C\, \dot C)] = 0$.
Furthermore, it can be checked that the CF-type restriction {\it also} remains invariant under a set of discrete symmetry transformations:
$B \rightarrow  - \bar B, \; \bar B \rightarrow  - B, \; C \rightarrow  \pm\, i\, \bar C,\; \bar C \rightarrow \pm\, i\, C$.

As claimed earlier, the {\it equivalence} of the {\it coupled} Lagrangians $L_B$ and $L_{\bar B}$ w.r.t. the off-shell nilpotent (anti-)BRST symmetries
 can be corroborated by the following explicit observations when we apply $s_b$ on $L_{\bar B}$ and $s_{ab}$ on  $L_{B}$, namely;
 \begin{eqnarray}
 s_{ab}\, L_{B}  & = & \frac {d}{d\tau}\Big [ \bar C\,L_f + e^2(\dot{\bar C}\,{\bar C}\, \dot C  - B \,  \dot {\bar C}) - B ^2\, \bar C +  e\,\dot e \,(\dot {\bar C} \,\bar C\, C - B \, \bar C)     
+ (2\,B - \bar{B})\dot {\bar C}\, \bar C\, C \Big]\nonumber\\
& + & (B + \bar B + \dot {\bar C}\, C - \bar C\, \dot C)\Big[2\,B\,\dot {\bar C} + 2\,\dot {\bar C}\,\bar C\,\dot C + \ddot{\bar C}\, \bar C \, C + e\,\dot e \, \dot {\bar C}\Big]\nonumber\\
& + & \frac{d}{d\tau}\Big [B +  {\bar B} + \dot {\bar C}\, C - \bar C\, \dot C\Big]\times  (B\, \bar C + e^2\, \dot {\bar C}),\nonumber\\
s_b\, L_{\bar B}  & = & \frac {d}{d\tau}\Big [ C\,L_f + e^2(\dot {\bar C}\, C\, \dot C + \bar B \,  \dot C) - \bar B ^2\, C +  e\, \dot e \,(\bar C\, C\, \dot C + \bar B \, C)  - 
(2\,\bar B - B)\, \bar C\, C\, \dot C  \Big]\nonumber\\
& + & (B + \bar B + \dot {\bar C}\, C - \bar C\, \dot C)\Big[2\,\bar B\,\dot C - 2\,\dot {\bar C}\,C\,\dot C +  \bar C\, \ddot C \, C - e\,\dot e \, \dot C \Big]\nonumber\\
& + &  \frac{d}{d\tau}\Big [B +  {\bar B} + \dot {\bar C}\, C
 - \bar C\, \dot C\Big] \times (\bar B\, C - e^2\, \dot C).
\end{eqnarray}
In other words, we note that {\it both} the Lagrangians respect {\it both} the 
nilpotent (anti-)BRST symmetries [cf. Eq. (47)] provided we take into account the validity of the CF-type 
restriction: $B + \bar{B} + (\dot{\bar{C}}\,C - \bar{C}\,\dot{C}) = 0$. Thus, it is crystal clear that the absolute anticommutativity
property  as well as the {\it equivalence} of the Lagrangians $L_B$ and $L_{\bar{B}}$ are true if and only 
if the CF-type restriction is taken into account. It is also evident that, under the validity of the {\it latter},
we have the following explicit expressions for symmetry transformations 
\begin{eqnarray}
s_{ab}\,L_B &=& \frac{d}{d\tau}\Big[\bar{C}\,L_f + e^2(\dot{\bar{C}}\,\bar{C}\,\dot{C} - B\,\dot{\bar{C}}) - B^2\,\bar{C}\nonumber\\
& + & e\,\dot{e}\,(\dot{\bar{C}}\,\bar{C}\,C - B\,\bar{C}) + (2\,B - \bar{B})\dot{\bar{C}}\,\bar{C}\,C \Big], \nonumber\\
s_b\,L_{\bar{B}} &=& \frac {d}{d\tau}\Big [ C\,L_f + e^2(\dot {\bar C}\, C\, \dot C + \bar B \,  \dot C) - \bar B ^2\, C, \nonumber\\
& + & e\,\dot e\,(\bar C\, C\, \dot C + \bar B \, C)  - 
(2\,\bar B - B)\, \bar C\, C\, \dot C \Big],
\end{eqnarray}
which render the action integrals $S_1 = \int{d\,\tau\,L_B}$ and $S_2 = \int{d\,\tau\,L_{\bar{B}}}$ of our theory 
(anti-)BRST invariant  for the physically well-defined variables that vanish-off as $\tau \longrightarrow  \pm \,\infty$
 when our theory is restricted to respect the (anti-)BRST invariant  CF-type restriction: 
$ B + \bar B + \dot{\bar C}\,C - \bar C\, \dot C = 0$.

According to the basic concepts behind the Noether theorem, the above continuous symmetries [i.e. (anti-)BRST symmetries] lead to the derivation of 
conserved and nilpotent (anti-)BRST charges. The {\it equivalent} expressions, for the conserved BRST charge, are
\begin{eqnarray}
Q_b ^{(1)} &=& B\,\bar C\,C\,\dot C - B^2\, C - B\,e^2\,\dot C - B\,e\,\dot e\, C +  \frac {1}{2}\,e\,C\,(p^2 - m^2),\nonumber\\
Q_b ^{(2)} &=& e^2\,(\dot B\,C - B\,\dot C  + \dot {\bar C}\,C\,\dot C) +  B\,\bar C\,C\,\dot C - B^2\, C -   B\,e\,\dot e\, C,  \nonumber\\
Q_b ^{(3)} &=&  e^2\,(\dot B\,C - B\,\dot C + \dot {\bar C}\,C\,\dot C),\nonumber\\
Q_b ^{(4)} &=&  e^2\,(\dot B\,C - B\,\dot C  + \dot {\bar C}\,C\,\dot C) + e^2\,\bar C\, C\, \ddot C +   2\,e\,\dot e\,\bar C\,C\,\dot C,\nonumber\\
&\equiv & s_b [e^2\,(\dot {\bar C}\,C - \bar C\, \dot C)],\nonumber\\
Q_b ^{(5)} &=& e^2\,(\bar{B}\,\dot C  - \dot{\bar{B}}\,C + 2\,\dot {\bar C}\,C\,\dot C) +2\,e\,\dot{e}\,\bar{C}\,C\,\dot{C},\nonumber\\
&\equiv & s_{ab}(e^2\,C\,\dot{C}).
\end{eqnarray}
Similarly, the {\it equivalent} forms of the conserved anti-BRST charge are:
\begin{eqnarray}
Q_{ab}^{(1)} &=& \bar{B}\,\bar C\,\dot{\bar{C}}\,C - \bar{B}^2\, \bar{C} + \bar{B}\,e^2\,\dot{\bar{C}} + \bar{B}\,e\,\dot e\,\bar{C}  +  \frac {1}{2}\,e\,\bar{C}\,(p^2 - m^2),\nonumber\\
Q_{ab} ^{(2)} &=& e^2\,(\bar{B}\,\dot{\bar{C}} - \dot{\bar{B}}\,\bar{C} + \dot {\bar C}\,\bar{C}\,\dot C) +  \bar{B}\,\bar C\,\dot{\bar{C}}\,C - \bar{B}^2\, \bar{C} +   \bar{B}\,e\,\dot e\, \bar{C},  \nonumber\\
Q_{ab} ^{(3)} &=& e^2\,(\bar{B}\,\dot{\bar{C}} - \dot{\bar{B}}\,\bar{C} + \dot {\bar C}\,\bar{C}\,\dot C),\nonumber\\
Q_{ab} ^{(4)} &=&  e^2\,(\bar{B}\,\dot{\bar{C}} - \dot{\bar{B}}\,\bar{C}+ \dot {\bar C}\,\bar{C}\,\dot C)- e^2\, \bar{C}\, \ddot{\bar{C}}\,C -   2\,e\,\dot e\,\bar{C}\,\dot{\bar{C}}\,C,\nonumber\\
&\equiv & s_{ab} [e^2\,(\bar{C}\,\dot {C} - \dot{\bar{C}}\,{C})], \nonumber\\
Q_{ab} ^{(5)} &=& e^2\,(\dot{B}\,\bar{C} - B\,\dot{\bar{C}} + 2\,\dot {\bar{C}}\,\bar{C}\,\dot{C}) +2\,e\,\dot{e}\,\dot{\bar{C}}\,\bar{C}\,C,\nonumber\\
&\equiv & s_{b}(e^2\,\dot{\bar{C}}\,\bar{C}).
\end{eqnarray}
The conservation law (i.e. $\dot Q_{(a)b} ^{(r)} = 0, r = 1, 2, 3, 4, 5$) can be proven in a straightforward manner by using the 
equations of motion derived from the Lagrangians $L_B$ and $L_{\bar B}$ [cf. Eqs. (53), (54) below].
We would like to point out  that the expressions $Q_b ^{(1)}$ and $Q_{ab} ^{(1)}$ have been  derived by
using the {\it direct} mathematical form of the Noether theorem.
However, the {\it other} equivalent forms for the charges  
have been  obtained  by using the equations of motion (EOM) for the Lagrangians $L_B$ and $L_{\bar{B}}$. In fact, the precise  forms of EOM from $L_B$ are:
\begin{eqnarray}
&& \dot{p_\mu} = 0,\quad \dot{x_\mu} = e\,p_\mu, \quad B + 2\,\dot{\bar{C}}\,C + e\,\dot{e} + \bar{C}\,\dot{C} = 0, \nonumber \\
&& e\,\dot{B} - e\,\dot{\bar{C}}\,\dot{C} + e\,\ddot{\bar{C}}\,C - \frac{1}{2}(p^2 - m^2) = 0, \nonumber\\ 
&&\dot{B}\,\bar{C} - B\,\dot{\bar{C}} + e\,\dot{e}\,\dot{\bar{C}} + e^2\,\ddot{\bar{C}} 
+ \bar{C}\,\ddot{\bar{C}}\,C + 2\,\bar{C}\,\dot{\bar{C}}\,\dot{C} = 0, \nonumber \\ 
&&- B\,\dot{C} - 2\,\dot{B}\,C - 3\,e\,\dot{e}\,\dot{C} - e^2\,\ddot{C} - e\,\ddot{e}\,C 
- \dot{e}^{2}\,C  + 2\,\dot{\bar{C}}\,C\,\dot{C} + \bar{C}\,C\,\ddot{C} = 0.
\end{eqnarray}
In exactly similar fashion, the {\it exact} forms of EOM from $L_{\bar{B}}$ are:
\begin{eqnarray}
 &&\dot{p_\mu} = 0,\quad \dot{x_\mu} = e\,p_\mu, \quad \bar{B} - 2\,{\bar{C}}\,\dot{C} - e\,\dot{e} - \dot{\bar{C}}\,{C} = 0,  \nonumber\\ 
&&e\,\dot{\bar{B}} - e\,{\bar{C}}\,\ddot{C} + e\,\dot{\bar{C}}\,\dot{C} + \frac{1}{2}(p^2 - m^2) = 0, \nonumber \\
&&\dot{\bar{B}}\,{C} - \bar{B}\,\dot{{C}} - e\,\dot{e}\,\dot{{C}} - e^2\,\ddot{{C}} + \bar{C}\,C\,\ddot{{C}}
 + 2\,\dot{\bar{C}}\,C\,\dot{{C}} = 0, \nonumber \\ 
&&- 2\,\dot{\bar{B}}\,\bar{C} - \bar{B}\,\dot{\bar{C}} + 3\,e\,\dot{e}\,\dot{\bar{C}} + e^2\,\ddot{\bar{C}} 
+ e\,\ddot{e}\,\bar{C} + \dot{e}^{2}\,\bar{C} + {\bar{C}}\,\ddot{\bar{C}}\,{C}  + 2\,\bar{C}\,\dot{\bar{C}}\,\dot{C} = 0. 
\end{eqnarray}
The above EOMs (53) and (54) can be used, in a straightforward fashion, to prove that {\it all} the 
(anti-)BRST charges, listed in (52) and (51), are conserved (i.e. $\dot{Q}_{(a)b}^{(r)} = 0, r = 1, 2, ..., 5$), primarily,
due to the basic concepts behind  Noether's theorem.

We have expressed the conserved and off-shell nilpotent (anti-)BRST charges in various forms [cf. Eqs. (52), (51)] because all 
the forms have their own significance. For instance, a close look at the $Q_{(a)b}^{(4)}$ establishes the 
nilpotency of the charges as it can be seen that:
\begin{eqnarray}
s_b\,Q_b^{(4)} = -i\,\{Q_b^{(4)}, Q_b^{(4)}\} = 0 & \Longrightarrow & (Q_b^{(4)})^2 = 0  \Longleftrightarrow   s_b^{2} = 0, \nonumber\\
s_{ab}\,Q_{ab}^{(4)} = -i\,\{Q_{ab}^{(4)}, Q_{ab}^{(4)}\} = 0 & \Longrightarrow & (Q_{ab}^{(4)})^2 = 0 \Longleftrightarrow   s_{ab}^{2} = 0.
\end{eqnarray}  
Thus, it is crystal clear (from the above equation) that the nilpotency of the (anti-)BRST symmetries 
are very intimately connected with the off-shell nilpotency of the (anti-)BRST charges. The expressions for 
the {\it equivalent} conserved (anti-)BRST charges $Q_{(a)b}^{(2, 3)}$ are the intermediate steps for
 obtaining the BRST {\it exact} form
of $Q_b^{(4)}$ and anti-BRST {\it exact} form of $Q_{ab}^{(4)}$. Furthermore, we would like to mention that the 
expressions for the conserved charges $Q_{(a)b}^{(5)}$ have been obtained from $Q_{(a)b}^{(4)}$ by using the 
beauty and strength of the CF-type restriction: $B + \bar B + \dot{\bar C}\,C - \bar C\, \dot C = 0$. The
expressions for the conserved (anti-)BRST charges $Q_{(a)b}^{(5)}$ are very interesting for us because they
encode in themselves the absolute anticommutativity property  
\begin{eqnarray}
s_{ab}\,Q_{b}^{(5)} &=& -i\,\{Q_{b}^{(5)}, Q_{ab}^{(5)}\} = 0 \quad \Longleftrightarrow  \quad s_{ab}^{2} = 0, \nonumber\\
s_b\,Q_{ab}^{(5)} &=& -i\,\{Q_{ab}^{(5)}, Q_b^{(5)}\} = 0 \quad \Longleftrightarrow  \quad s_b^{2} = 0, 
\end{eqnarray}
where we have used the basic principle behind the connection between the continuous symmetry transformations
$s_{(a)b}$ and their generators as the conserved Noether (anti-)BRST charges. We would like to lay emphasis on the 
fact that it is the power and potential of the CF-type restriction that has enabled us to express the conserved BRST 
charge $(Q_b^{(5)})$ as an anti-BRST {\it exact} quantity and the conserved anti-BRST charge $(Q_{ab}^{(5)})$ 
as the BRST {\it exact} object. In our Appendix B, we discuss more about the absolute anticommutativity of the nilpotent 
(anti-)BRST conserved charges and the existence of  (anti-)BRST invariant CF-type restriction on our theory.

In some sense, the above exercise is a reflection of our observations  in Eq. (48) where we have shown that 
the absolute anticommutativity property $(s_b\,s_{ab} + s_{ab}\,s_b = 0)$ of the (anti-)BRST symmetries 
$s_{(a)b}$ are {\it true} {\it only} on a submanifold, in the space of {\it quantum} variables, which is defined by the 
CF-type equation: $B + \bar B + \dot{\bar C}\,C - \bar C\, \dot C = 0$. Since, the nilpotency and absolute 
anticommutativity properties are very {\it sacrosanct} in the BRST formalism, the requirement of the 
{\it latter} property for the conserved charges, in our present discussion, leads to the derivation 
of the CF-type restriction (24) which was {\it also} derived from the {\it modified} BT-supervariable 
approach (MBTSA) to BRST formalism (cf. Sec. 3). In other words, we take {\it directly} the help of the CF-type 
restriction to recast the conserved (anti-)BRST charges in a {\it specific} form [e.g. $(Q_{(a)b}^{(5)})$] such 
that the BRST charge is expressed as an anti-BRST {\it exact} quantity (and the anti-BRST charge as 
the BRST {\it exact} form). At this juncture, it is crystal clear that the absolute anticommutativity
of (i) the nilpotent (anti-)BRST symmetries [cf. Eq. (48)], and (ii) the conserved and nilpotent
(anti-)BRST charges [cf. Eq. (56)] owe their 
origin to the CF-type restriction: $B + \bar B + \dot{\bar C}\,C - \bar C\, \dot C = 0$ 
on our 1D diffeomorphism invariant scalar relativistic particle.

\section{Invariance of the Lagrangians, Nilpotency and Anticommutativity of the Conserved  Charges: ACSA}

\noindent
We now capture the (anti-)BRST invariance of the coupled Lagrangians within the framework of ACSA to
BRST formalism and thereby prove the existence of the CF-type restriction (24) on our theory from 
the point of view of the {\it symmetry} considerations\footnote{To be precise, we actually capture the
(anti-)BRST invariance  of the coupled (but equivalent) Lagrangians $L_B$ and $L_{\bar B}$ [cf. Eq. (46)].
Furthermore, we also describe our observations in equation (49) in the language of the ACSA which establishes the existence of the CF-type restriction: $B + \bar B + \dot{\bar C}\,C - \bar C\, \dot C = 0$
 on our theory in terms of the invariance of the action integrals.}. In this context,  first of all, we generalize the BRST invariant Lagrangian 
$L_B$ to its counterpart super Lagrangian $\tilde L_B^{(ac)}$ on the (1, 1)-dimensional
{\it anti-chiral} super sub-manifold of the {\it general} (1, 2)-dimensional supermanifold (on which
our theory is generalized) as follows
\begin{eqnarray*}
L_B  &\longrightarrow &   \tilde L_B^{(ac)}  =   P^{(ha)}_\mu (\tau, \bar\theta)\,{\dot X}^{\mu(ha)} (\tau, \bar\theta) \nonumber\\  
& - & \frac {1}{2}\,E^{(b)}(\tau, \bar\theta)\,\Big[P^{(ha)}_\mu (\tau, \bar\theta) \,P^{\mu(ha)} (\tau, \bar\theta) - m^2\Big]
\nonumber\\ 
& - &\frac {1}{2}\,{\tilde B}^{(b)} (\tau, \bar\theta)\,{\tilde B}^{(b)} (\tau, \bar\theta) -   {\tilde B}^{(b)} (\tau, \bar\theta)\, \Big[E^{(b)}(\tau, \bar\theta)\,{\dot E}^{(b)} (\tau, \bar\theta) \nonumber\\
& + &2\,{\dot{\bar F}}^{(b)} (\tau, \bar\theta) \,F^{(b)}(\tau, \bar\theta) + {\bar F}^{(b)}(\tau, \bar\theta)\,{\dot F}^{(b)} (\tau, \bar\theta)\Big]
 \nonumber \\
\end{eqnarray*}
\begin{eqnarray}  
& - &E^{(b)}(\tau, \bar\theta)\,E^{(b)}(\tau, \bar\theta)\,{\dot{\bar F}}^{(b)} (\tau, \bar\theta)\,{\dot F}^{(b)} (\tau, \bar\theta) \nonumber \\ 
& - &E^{(b)}(\tau, \bar\theta)\,{\dot E}^{(b)} (\tau, \bar\theta)\,{\dot{\bar F}}^{(b)} (\tau, \bar\theta)\,F^{(b)} (\tau, \bar\theta) \nonumber \\ 
& - &{\bar F}^{(b)} (\tau, \bar\theta)\,{\dot{\bar F}}^{(b)} (\tau, \bar\theta)\,F^{(b)}(\tau, \bar\theta)\,{\dot F}^{(b)} (\tau, \bar\theta),
\end{eqnarray}
where it can be noted that we have $\tilde B^{(b)} (\tau, \bar\theta) = B (\tau)$ because of the fact that $s_b \, B = 0$. Thus, even though,
we have written $\tilde B^{(b)} (\tau, \bar\theta)$ in the above  equation (57), it is actually an {\it ordinary} Nakanishi-Lautrup type auxiliary variable
$B (\tau)$ of our theory [cf. Eq. (45)]. Now we are in the position  to capture the BRST invariance of the Lagrangian $L_B$ [cf. Eq. (46)]
in the language of ACSA as: 
\begin{eqnarray}
\frac{\partial} {\partial \bar\theta }\, \tilde L_B^{(ac)} & = & \frac {d}{d\tau}\Big [C\,L_f - e^2\,B\,\dot C 
- e\,\dot e\,B\, C -  B\,\bar C\,\dot C\, C -  B^2\,C\Big] \equiv   s_b\, L_B.
\end{eqnarray}
Geometrically, it implies that the {\it anti-chiral} super Lagrangian $\tilde L_B^{(ac)} $ is a combination of the suitable (super)variables such that
its translation along the $\bar\theta$-direction of the (1, 1)-dimensional {\it anti-chiral} super sub-manifold produces a total ``time'' derivative
in the {\it ordinary} space thereby rendering the action integral, in the {\it ordinary} space, invariant under the BRST symmetry transformations $s_b$
due to the Gauss divergence theorem. It should be noted that the BRST transformations ($s_b$)
is identified with the translational generator $\partial_{\bar\theta}$  [9-11] on the {\it anti-chiral} super sub-manifold  [of 
the {\it general} (1, 2)-dimensional
supermanifold on which our 1D system of a scalar non-supersymmetric  free relativistic particle is generalized].

We now discuss the anti-BRST invariance of our theory.
To capture the anti-BRST symmetry invariance [cf. Eq. (46)] of the Lagrangian
 $L_{\bar B}$, first of all, we generalize the {\it latter} to the
(1, 1)-dimensional {\it chiral} super sub-manifold [of the general (1, 2)-dimensional
supermanifold on which our system of a 1D ordinary free non-supersymmetric  scalar relativistic particle is considered] as 
\begin{eqnarray}
L_{\bar B} & \longrightarrow &   \tilde L_{\bar B} ^{(c)}  =  P^{(hc)}_\mu (\tau, \theta)\,{\dot X}^{\mu(hc)}
(\tau, \theta)\nonumber\\ 
& - & \frac {1}{2}\,E^{(ab)}(\tau, \theta)\,\big[P^{(hc)}_\mu (\tau, \theta) \,P^{\mu(hc)} (\tau, \theta) - m^2\big] \nonumber \\
& - & \frac {1}{2}\,{\tilde{\bar B}}^{(ab)}(\tau, \theta)\,{\tilde{\bar B}}^{(ab)}(\tau, \theta) +   {\tilde{\bar B}}^{(ab)}(\tau, \theta)\, \big[E^{(ab)}(\tau, \theta)\,{\dot E}^{(ab)}(\tau, \theta) \nonumber \\ 
& + & 2\,{\bar F}^{(ab)}(\tau, \theta) \,{\dot F}^{(ab)}(\tau, \theta) + {\dot{\bar F}}^{(ab)}(\tau, \theta)\, F^{(ab)}(\tau, \theta)\big] \nonumber\\
 & - & E^{(ab)}(\tau, \theta)\,E^{(ab)}(\tau, \theta)\,{\dot{\bar F}}^{(ab)}(\tau, \theta)\,{\dot F^{(ab)}}(\tau, \theta) \nonumber \\
 & - & E^{(ab)}(\tau, \theta)\,{\dot E^{(ab)}}(\tau, \theta)\,{\bar F}^{(ab)}(\tau, \theta)\,{\dot F^{(ab)}}(\tau, \theta) \nonumber\\
 & - & {\bar F}^{(ab)}(\tau, \theta)\,{\dot{\bar F}}^{(ab)}(\tau, \theta)\,F^{(ab)}(\tau, \theta)\,{\dot F^{(ab)}}(\tau, \theta),
\end{eqnarray}
where it can be noted that $\tilde {\bar B} ^{(ab)}(\tau, \theta) = \bar B (\tau)$ because of the 
fact that $s_{ab} \;\bar B = 0$. Thus, even though, we have written
${\tilde{\bar B}}^{(ab)}(\tau, \theta)$ in our {\it chiral} super Lagrangian $\tilde L_{\bar B} ^{(c)}$, it is actually an 
ordinary variable $\bar B (\tau) $. The anti-BRST invariance of the above chiral super Lagrangian
 can be expressed, in terms of the translational generator along 
the Grassmannian $\theta$-direction (with the input $\partial_\theta \longleftrightarrow s_{ab}$), as:
\begin{eqnarray}
\frac{\partial} {\partial \theta }\, \tilde L_{\bar B}^{(c)} & = & \frac {d}{d\tau}\Big [ \bar C\,L_f + e^2\,\bar B\,\dot {\bar C} 
+ e\,\dot e\,\bar B\, \bar C  -  \bar B\, \dot {\bar C} \,\bar C\,C -{\bar B}^2\,\bar C\Big] \equiv  \, s_{ab}\, L_{\bar B},
\end{eqnarray}
where $\partial_\theta$ is the translational generator [9-11] along the 
Grassmannian (i.e. $\theta$) direction of the (1, 1)-dimensional
{\it chiral} super sub-manifold of the general (1, 2)-dimensional supermanifold. 
Once again, we note that, geometrically, the {\it chiral} super Lagrangian
 $\tilde L_{\bar B} ^{(c)}$ is a specific combination of the appropriate 
chiral (super)variables such that its translation along the $\theta$-direction of the {\it chiral}
 super sub-manifold generates a total ``time'' derivative (in the {\it ordinary} space) 
thereby rendering the action integral (in the ordinary
 space) invariant under the anti-BRST symmetry transformations ($s_{ab}$) due to the Gauss divergence theorem. In the language of
 ACSA to BRST formalism, we note that the super action integral $\tilde S_1 = 
\int d \theta \int_{-\infty }^{+\infty } d \tau \; \tilde L_{\bar B} ^{(c)}$ 
 remains invariant under the anti-BRST transformations.

We establish now the existence of the CF-type restriction: $B + \bar B + \dot{\bar C}\,C
 - \bar C\, \dot C = 0$ within the framework of the
ACSA to BRST formalism by considering the anti-BRST invariance of the a {\it chiral}
 super Lagrangian  $\tilde L_B^{(c)} $ and BRST invariance of the 
{\it anti-chiral} super Lagrangian $\tilde L_{\bar B} ^{(ac)}$. This is due to 
the fact that, as claimed in our earlier discussions (cf. Sec. 5),
both the Lagrangians $L_B$ and $L_{\bar B}$ respect {\it both} the symmetries
 provided the theory is considered on a sub-manifold of the
space of {\it quantum} variables where  the CF-type restriction is satisfied.  
Towards this goal in mind, we note the following generalization:
\begin{eqnarray}
L_{\bar B} & \longrightarrow & \tilde L_{\bar B} ^{(ac)}  =   P^{(ha)}_\mu (\tau, \bar\theta)\,{\dot X}^{\mu(ha)} (\tau, \bar\theta) \nonumber\\ 
& - & \frac {1}{2}\,E^{(b)}(\tau, \bar\theta)\,\Big[P^{(ha)}_\mu (\tau, \bar\theta) \,P^{\mu(ha)} (\tau, \bar\theta) - m^2\Big] \nonumber \\
& - &\frac {1}{2}\,{\tilde{\bar B}}^{(b)}(\tau, \bar\theta)\,{\tilde{\bar B}}^{(b)}(\tau, \bar\theta) +  {\tilde{\bar B}}^{(b)}(\tau, \bar\theta)\, \Big[E^{(b)}(\tau, \bar\theta)\,{\dot E}^{(b)}(\tau, \bar\theta) \nonumber\\ 
& + & 2\,{\bar F}^{(b)}(\tau, \bar\theta) \,{\dot F}^{(b)}(\tau, \bar\theta) + {\dot{\bar F}}^{(b)}(\tau, \bar\theta)\,F^{(b)}(\tau, \bar\theta)\Big] \nonumber \\
 & - & E^{(b)}(\tau, \bar\theta)\,E^{(b)}(\tau, \bar\theta)\,{\dot{\bar F}}^{(b)}(\tau, \bar\theta)\,{\dot F}^{(b)}(\tau, \bar\theta) \nonumber \\
 & - & E^{(b)}(\tau, \bar\theta)\,{\dot E}^{(b)}(\tau, \bar\theta)\,{\bar F}^{(b)}(\tau, \bar\theta)\,{\dot F}^{(b)}(\tau, \bar\theta)\nonumber\\
 & - & {\bar F}^{(b)}(\tau, \bar\theta)\,{\dot{\bar F}}^{(b)}(\tau, \bar\theta)\,F^{(b)}(\tau, \bar\theta)\,{\dot F}^{(b)}(\tau, \bar\theta).
\end{eqnarray}
In the above, it should be noted that we have generalized the {\it perfectly} 
 anti-BRST invariant Lagrangian $L_{\bar B}$ to its counterpart {\it anti-chiral} 
super Lagrangian $\tilde L_{\bar B}^{(ac)} $ on the
(1, 1)-dimensional {\it anti-chiral} super sub-manifold [of the general (1, 2)-dimensional
 supermanifold]. We are in the position now to apply
a derivative $\partial_{\bar\theta}$ w.r.t. $\bar \theta$ on the above super 
Lagrangian which yields the following (with the input: 
$s_b \leftrightarrow \partial_{\bar\theta})$:
\begin{eqnarray}
\frac{\partial} {\partial \bar \theta}\; \tilde L_{\bar B} ^{(ac)}  & = & 
\frac {d}{d\tau}\Big [ C\,L_f + e^2(\dot {\bar C}\, C\, \dot C + \bar B \,  \dot C)  +    e\,\dot e(\bar C\, C\, \dot C + \bar B \, C)\nonumber\\ 
&  - &  (2\,\bar B - B)\, \bar C\, C\, \dot C -  \bar B ^2\, C\Big] +   (B + \bar B + \dot {\bar C}\, C - \bar C\, \dot C)\nonumber\\ 
&\times &(2\,\bar B\,\dot C  -  2\,\dot {\bar C}\,C\,\dot C 
+  \bar C\, \ddot C \, C - e\,\dot e \, \dot C)\nonumber\\
& + &  \frac{d}{d\tau}\Big [B +  {\bar B} + \dot {\bar C}\, C - \bar C\, \dot C\Big](\bar B\, C - e^2\, \dot C) \;\equiv \; s_b\, L_{\bar B}.
\end{eqnarray}
The above equation leads to the derivation of the CF-type restriction in the sense that the   
{\it anti-chiral} super Lagrangian $\tilde L_{\bar B} ^{(ac)} $, when operated by
$ \partial_{\bar\theta}$, produces a total ``time'' derivative 
{\it plus} terms that vanish-off on the submanifolds of the space of variables which is defined by the CF-type restriction: 
$B +  {\bar B} + \dot {\bar C}\, C - \bar C\, \dot C$. With the identification:
 $s_b \leftrightarrow \partial_{\bar\theta}$, it is clear that we
have obtained the {\it same} relationship as given in equation (49)
 in the {\it ordinary} space for the BRST symmetry transformation of $L_{\bar B}$ (i.e. $s_b \,L_{\bar B}$).

In exactly similar fashion, we can generalize the {\it perfectly}  
BRST invariant Lagrangian $L_B$ to its counterpart {\it chiral} super Lagrangian $\tilde L_B^{(c)} $ as follows:
\begin{eqnarray}
L_{B} & \longrightarrow &   \tilde L_{B} ^{(c)}   =   P^{(hc)}_\mu (\tau, \theta)\,{\dot X}^{\mu(hc)} (\tau, \theta)\nonumber\\  
& - & \frac {1}{2}\,E^{(ab)}(\tau, \theta)\,\big[P^{(hc)}_\mu (\tau, \theta) \,P^{\mu(hc)} (\tau, \theta) - m^2\big] \nonumber \\
& - &\frac {1}{2}\,{\tilde{ B}}^{(ab)}(\tau, \theta)\,{\tilde{B}}^{(ab)}(\tau, \theta) -  {\tilde{ B}}^{(ab)}(\tau, \theta)\, \big[E^{(ab)}(\tau, \theta)\,{\dot E}^{(ab)}(\tau, \theta) \nonumber \\ 
& + &2\, {\dot {\bar F}}^{(ab)}(\tau, \theta) \,{ F}^{(ab)}(\tau, \theta) 
+ {\bar F}^{(ab)}(\tau, \theta)\,{\dot F}^{(ab)}(\tau, \theta)\big] \nonumber\\
 & - &E^{(ab)}(\tau, \theta)\,E^{(ab)}(\tau, \theta)\,{\dot{\bar F}}^{(ab)}(\tau, \theta)\,{\dot F}^{(ab)}(\tau, \theta) \nonumber \\
 & - &E^{(ab)}(\tau, \theta)\,{\dot E}^{(ab)}(\tau, \theta)\, {\dot {\bar F}}^{(ab)}(\tau, \theta)\,{F}^{(ab)}(\tau, \theta) \nonumber\\
 & - & {\bar F}^{(ab)}(\tau, \theta)\,{\dot{\bar F}}^{(ab)}(\tau, \theta)\,F^{(ab)}(\tau, \theta)\,{\dot F}^{(ab)}(\tau, \theta),
\end{eqnarray}
where all the symbols and notations have been clarified earlier. At this juncture, we apply a Grassmannian derivative $\partial_\theta$ on the above 
super Lagrangian which yields the following:
\begin{eqnarray}
\frac{\partial} {\partial \theta}\; \tilde L_{B} ^{(c)}  & = & 
 \frac {d}{d\tau}\Big [ \bar C\,L_f + e^2(\dot{\bar C}\,{\bar C}\, \dot C 
 - B \,  \dot {\bar C}) +   e\,\dot e(\dot {\bar C} \,\bar C\, C - B \, \bar C)\nonumber\\      
 & + &  (2\,B - \bar{B})\dot {\bar C}\, \bar C\, C - B ^2\, \bar C\Big] +  (B + \bar B + \dot {\bar C}\, C - \bar C\, \dot C)\nonumber\\ 
&\times &(2\,B\,\dot {\bar C} + 2\,\dot {\bar C}\,\bar C\,\dot C +  \ddot{\bar C}\, \bar C \, C + e\,\dot e \, \dot {\bar C})\nonumber\\
& + & \frac{d}{d\tau}\Big [B +  {\bar B} + \dot {\bar C}\, C - \bar C\, \dot C\Big] (B\, \bar C + e^2\, \dot {\bar C})\;\equiv  \; s_{ab}\, L_{B}.
\end{eqnarray}
Thus, we note that we have derived the observation that has been made in equation (49). In other words,  the ACSA to BRST formalism leads to the derivation
of the CF-type restriction when we consider the anti-BRST invariance of the  
{\it perfectly} BRST invariant Lagrangian $L_B$ as well as the BRST invariance  of
the {\it perfectly} anti-BRST invariant Lagrangian $L_{\bar B}$  of our theory.

At this stage, we would like to capture the off-shell nilpotency as well as the 
absolute anticommutativity of the {\it conserved} (anti-)BRST charges
[cf. Eqs. (55), (56)] within the framework of the ACSA to BRST formalism. 
Towards this goal in mind, we note that, out of the equivalent expressions for the conserved 
(anti-) BRST charges quoted in (52) and (51), one set of the conserved charges
 $Q_{(a)b}^{(4)}$ have been expressed in the (anti-)BRST {\it exact} forms.
Keeping in mind the identifications: $s_b \leftrightarrow \partial_{\bar\theta},
 s_{ab} \leftrightarrow \partial_\theta $, we note the following:
\begin{eqnarray*}
Q_b ^{(4)} & = & \frac{\partial}{\partial\bar{\theta}}\Big[E^{(b)}(\tau, \bar{\theta})\,E^{(b)}(\tau, \bar{\theta})\big\{{\dot{\bar{F}}}^{(b)}(\tau, \bar{\theta})\,F^{(b)}(\tau, \bar{\theta})-   {\bar{F}}^{(b)}(\tau, \bar{\theta})\,{\dot{F}}^{(b)}(\tau, \bar{\theta})\big\}\Big], \nonumber\\
\end{eqnarray*}
\begin{eqnarray} 
&\equiv & \int d\bar\theta\,\Big[E^{(b)}(\tau, \bar{\theta})\,E^{(b)}(\tau, \bar{\theta})\big\{{\dot{\bar{F}}}^{(b)}(\tau, \bar{\theta})\,F^{(b)}(\tau, \bar{\theta})-   {\bar{F}}^{(b)}(\tau, \bar{\theta})\,{\dot{F}}^{(b)}(\tau, \bar{\theta})\big\}\Big], \nonumber\\
Q_{ab} ^{(4)} & = & \frac{\partial}{\partial\theta}\Big[E^{(ab)}(\tau, \theta)\,E^{(ab)}(\tau, \theta)\big\{{\bar{F}}^{(ab)}(\tau, \theta)\,{\dot{F}}^{(ab)}(\tau, \theta) -   {\dot{\bar{F}}}^{(ab)}(\tau, \theta)\,F^{(ab)}(\tau, \theta)\big\}\Big], \nonumber\\
&\equiv & \int d\theta  \Big[E^{(ab)}(\tau, \theta) E^{(ab)}(\tau, \theta)\big\{{\bar{F}}^{(ab)}(\tau, \theta) {\dot{F}}^{(ab)}(\tau, \theta) - {\dot{\bar{F}}}^{(ab)}(\tau, \theta) F^{(ab)}(\tau, \theta)\big\}\Big].
\end{eqnarray}
As a consequence, it is straightforward to point out the  fact that we have the validity of the following relationships:
\begin{eqnarray}
\partial_\theta \; Q^{(4)}_{ab} &=& 0 \quad \Longleftrightarrow \quad \partial_\theta^2 = 0 \quad \Longleftrightarrow \quad s_{ab}^2 = 0, \nonumber\\
\partial_{\bar\theta} \; Q^{(4)}_{b} &=& 0 \quad \Longleftrightarrow \quad  \partial_{\bar\theta}^2 = 0 \quad \Longleftrightarrow \quad \, s_{b}^2 = 0.
\end{eqnarray}
In other words, we note that the nilpotency (i.e. $\partial_\theta^2 = 0, \partial_{\bar\theta}^2 = 0$) of the translational generators 
($\partial_\theta, \partial_{\bar\theta}$) along the $(\theta)\bar\theta$-directions of the (1, 1)-dimensional {\it chiral} and {\it anti-chiral}
 super sub-manifolds
[of the {\it general} (1, 2)-dimensional supermanifold] are responsible for 
capturing the off-shell nilpotency of the conserved (anti-)BRST charges $Q_{(a)b}^{(4)}$.
To be more precise, we further point out that the off-shell nilpotency of the 
conserved BRST charge $Q_{b}^{(4)}$ is connected with the nilpotency 
(i.e.  $\partial_{\bar\theta}^2 = 0$) of the translational generator  $\partial_{\bar\theta}$ along the $\bar\theta$-direction of the (1, 1)-dimensional
{\it anti-chiral} super submanifold. However, the off-shell nilpotency of the 
conserved anti-BRST charge $Q_{ab}^{(4)}$ is intimately connected with the nilpotency
(i.e.  $\partial_{\theta}^2 = 0$) of the translational generator $\partial_{\theta}$ along the $\theta$-direction of the (1, 1)-dimensional
{\it chiral} super submanifold of the {\it general} (1, 2)-dimensional supermanifold.

We concentrate, finally, on the proof of the absolute anticommutativity [cf. Eq. (56)] of the conserved  and nilpotent 
(anti-)BRST charges within the framework of the ACSA
to BRST formalism. In this context, we note that, from the list of the equivalent forms of the conserved (anti-)BRST charges, the BRST charge
$Q_{b}^{(5)}$ has been expressed as the anti-BRST {\it exact} quantity. On the other hand, the conserved anti-BRST charge $Q_{ab}^{(5)}$ has been
written in the BRST {\it exact} form. With the identifications: $s_b \leftrightarrow \partial_{\bar\theta}, s_{ab} \leftrightarrow \partial_\theta $, 
we  observe the sanctity of the following:
\begin{eqnarray}
&&Q_b ^{(5)} =  \frac{\partial}{\partial\theta}
\Big[E^{(ab)}(\tau,\theta)\,E^{(ab)}(\tau,\theta)\,F^{(ab)}(\tau,\theta)\,{\dot{F}}^{(ab)}(\tau,\theta)\Big], 
\nonumber \\
&& \equiv  \int d\theta\,\Big[E^{(ab)}(\tau,\theta)\,E^{(ab)}(\tau,\theta)\,F^{(ab)}(\tau,\theta)\,{\dot{F}}^{(ab)}(\tau,\theta)\Big], \nonumber \\
&& Q_{ab} ^{(5)}  =  \frac{\partial}{\partial\bar{\theta}}\Big[E^{(b)}(\tau,\bar{\theta})\,E^{(b)}(\tau, \bar{\theta})\,{\dot{\bar{F}}}^{(b)}(\tau, \bar{\theta})\,{\bar{F}}^{(b)}(\tau,\bar{\theta})\Big],\nonumber\\
&& \equiv  \int d\bar\theta\,\Big[E^{(b)}(\tau,\bar{\theta})\, E^{(b)}(\tau, \bar{\theta})\,{\dot{\bar{F}}}^{(b)}(\tau, \bar{\theta})\,{\bar{F}}^{(b)}(\tau,\bar{\theta})\Big].
\end{eqnarray}
As a consequence, it is straightforward that the followings results are true, namely; 
\begin{eqnarray}
\partial_\theta \; Q^{(5)}_{b} &=& 0 \quad \Longleftrightarrow \quad \partial_\theta^2 = 0 \quad \Longleftrightarrow \quad s_{ab}^2 = 0, \nonumber\\
\partial_{\bar\theta} \; Q^{(5)}_{ab} &=& 0 \quad \Longleftrightarrow \quad  \partial_{\bar\theta}^2 = 0 \quad \Longleftrightarrow \quad s_{b}^2 = 0.
\end{eqnarray}
Thus, it is crystal clear that, in the {\it ordinary} space,  the above equation is equivalent to equation (56) where we have proven the 
absolute anticommutativity of the conserved and off-shell nilpotent
(anti-)BRST charges. In the terminology of the ACSA to BRST formalism, we note that the absolute anticommutativity of the BRST
charge {\it with} the anti-BRST charge is deeply connected with the nilpotency (i.e.  $\partial_{\theta}^2 = 0$) of the translational 
generator $\partial_\theta$ along the Grassmannian direction $\theta$ of the (1, 1)-dimensional {\it chiral} super sub-manifold of the general  (1, 2)-dimensional
supermanifold on which our 1D theory is generalized. This should be contrasted with our earlier observation of the off-shell nilpotency of the BRST charge
(within the framework of ACSA to BRST formalism)
where it is the nilpotency (i.e.  $\partial_{\bar\theta}^2 = 0$) of the translational generator  $\partial_{\bar\theta}$ along the Grassmannian direction
$ \bar\theta$ of the (1, 1)-dimensional {\it anti-chiral} super sub-manifold that plays a decisive role. Similar kinds of statements could be made for the
proof of the absolute anticommutativity of the anti-BRST charge {\it with} the BRST charge. However, for the sake of {\it brevity}, we do {\it not} wish to
make any statement, in this regard, at this juncture of our discussion.\\

\section{Conclusions}

In our present endeavor, we have applied the BT-superfield/supervariable approach [9-11] in its {\it modified} 
form where the infinitesimal diffeomorphism transformation has been consistently taken into account [20, 21].
First of all, we have generalized the 1D infinitesimal diffeomorphism (i.e. reparameterization) transformation:
$\tau \rightarrow {\tau}'  = \tau - \epsilon(\tau)$ to its counterpart {\it superspace} infinitesimal 
reparameterization [cf. Eq. (13)] on the (1, 2)-dimensional supermanifold where the (anti-)ghost 
variables $(\bar C)C$ appear as the coefficients of the Grassmannian variables. This superspace reparameterization 
transformation has been incorporated into the supervariables [defined on the (1, 2)-dimensional supermanifold] 
and, then {\it only}, the super expansions along all possible Grassmannian directions of the
 {\it above} supermanifold  have been taken into account
in our present endeavor. After that, we have applied the HC [cf. Eq. (18)] to obtain the {\it quantum} (anti-)BRST 
 transformations corresponding to the {\it classical} infinitesimal reparameterization  
transformation: $\tau \rightarrow {\tau}'  = \tau - \epsilon(\tau)$ of our 1D theory. We have christened {\it this} approach 
as the {\it modified} BT-supervariable/superfield approach (MBTSA) to BRST formalism [20, 21].

One of the highlights of our present investigation is the derivation of the CF-type restriction:
$B + \bar B + \dot{\bar C}\,C - \bar C\, \dot C = 0$ by exploiting the power and potential of 
the MBTSA which has {\it also} led to the 
derivation of the off-shell nilpotent (anti-)BRST symmetries for the target space variables. 
The (anti-)BRST symmetry transformations for the {\it other} variables of our theory 
have been derived by using the {\it newly} proposed ACSA to BRST formalism [15-18] where
the (anti-)BRST invariant restrictions on the supervariables have played
a decisive role. We have also provided the proof of the existence of the
CF-type restrictions on our theory by considering (i) the symmetry invariance of the 
coupled (but equivalent) Lagrangians in the {\it ordinary} space, (ii) the (anti-)BRST
invariance of the super Lagrangians by exploiting the  potential and  power of the ACSA
to BRST formalism in the {\it superspace}, and (iii) the requirement of the proof of the absolute anticommutativity 
of the conserved (anti-)BRST charges. We have established that
the absolute anticommutativity of the (anti-)BRST symmetries (as well as 
corresponding conserved charges) and {\it equivalence} of the coupled (but equivalent) Lagrangians owe their origin 
to the (anti-)BRST invariant CF-type restriction (cf. Appendix A below).

In our present endeavor, we have applied the MBTSA
to derive the {\it proper}  nilpotent (anti-)BRST symmetry transformations for the
 phase space variables $x_\mu(\tau)$ and $p_\mu (\tau)$ of the target space. 
Rest of the (anti-)BRST symmetries for the other variables of our theory have been derived by using the ACSA.
 One of the {\it novel} observations of our present endeavor is the proof of the off-shell nilpotency and absolute anticommutativity 
of the conserved (anti-)BRST charges within the framework of ACSA. In this context, one interesting result
is the observation that the absolute anticommutativity of the BRST charge {\it with} the anti-BRST charge is deeply connected with the 
nilpotency ($\partial_\theta ^2 = 0$) of the translational generator $\partial_\theta $ along the $\theta$-direction of the {\it chiral}
super sub-manifold of the general (1, 2)-dimensional supermanifold. However, the absolute anticommutativity of the anti-BRST charge {\it with} 
the BRST charge is intimately connected with the nilpotency ($\partial_{\bar\theta }^2 = 0$) of the Grassmannian translational generator 
$\partial_{\bar\theta}$ along the $\bar\theta$-direction of the {\it anti-chiral} super submanifold of the general (1, 2)-dimensional 
supermanifold. Thus, in some sense, the ACSA distinguishes between the {\it chiral} and {\it anti-chiral} super sub-manifolds
as far as the proof of the absolute anticommutativity property.

As a closing remark on our present investigation, we would like to lay emphasis  on the fact that the
BRST quantization of a 1D free scalar relativistic particle is now a standard text-book material [19]
 where the infinitesimal gauge symmetry transformations: $\delta_g x_\mu = \xi\,p_\mu, \; \delta_g p_\mu = 0,\;
\delta_g e = \dot \xi $ have been exploited (cf. Sec. 2 for details). The central theme of our present endevor
is, however, to perform the consistent BRST quantization of our system by exploiting the full {\it classical} infinitesimal 
reparameterization symmetry transformations (5). This has been accomplished in the hope that our present understanding 
would help us to go to the higher dimensional  diffeomorphism invariant theories [e.g. (super)strings and gravitational theories]
where our theoretical method would be useful. It is gratifying to mention, in this context, that the reparameterization invariant model of 
a 1D scalar relativistic particle has already been systematically generalized to its counterpart  bosonic string in [24].
Another interesting point which we would like to mention is the
observation that the nature and form of the CF-type restriction is {\it universal} 
in the sense that we have obtained the same {\it restriction}
in the BRST quantization of the 1D spinning (i.e. supersymmetric)
relativistic particle [25] and a non-relativistic (as well as non-supersymmetric)
 free particle [26].

We have  discussed the (anti-)BRST symmetries and BRST quantization of the
D-dimensional diffeomorphism invariant theories with scalars, contravariant as well as
covariant vectors and metric tensor as well as its inverse, etc.  
(see, e.g. [21]). This has enabled us to discuss the (anti-)BRST symmetries for the 
affine connection, curvature tensor, curvature scalar,  etc., for the gravitational theories.
 Our future plan  is to discuss the BRST quantization of the diffeomorphism 
invariant theories [e.g. (super)string theories, gravitational theories, etc.] which are very popular at 
the frontier level of research  in the domain of theoretical high energy physics. 
It is gratifying  that we have taken a modest step in this direction in our 
earlier and recent works [27, 28] for a 2D diffeomorphism invariant model of bosonic string and derived 
the CF-type restrictions as: $B^a + \bar B^a + i\, (C^b \,\partial_b \bar C^a  + \bar C^b \, \partial_b C^a) = 0$
(with $a, b =  0, 1)$ which are the 2D version of the {\it universal} CF-type restrictions 
for the D-dimensional diffeomorphism invariant theory [20, 21].\\

\section*{Acknowledgments}

The present investigation has been carried out under the DST-INSPIRE fellowship (Govt. of India) awarded to B. Chauhan as well as the
BHU-fellowships received  by  A. Tripathi  and A. K. Rao. All these authors express their deep  sense of gratefulness to
the above national and local level  funding agencies for the financial support.
The authors thankfully acknowledge fruitful  suggestions/comments by the Reviewer which have made the quality of 
presentation more beautiful and correct.\\

\begin{center}
{\bf Appendix A: On the (Anti-)BRST Invariance\\ of the CF-Type Restriction: ACSA}\\
\end{center}
One of the key consequences of the geometrical BT-superfield/supervariable approach [9-11] is the observation
that it leads to the derivation of the (anti-)BRST invariant CF-type restriction in the context
of gauge theories. This is {\it also} true when we apply the {\it modified} version of BT-supervariable
approach [20, 21] to our 1D reparameterization/diffeomorphism invariant theory. It is straightforward to 
check that the CF-type restriction: $B + \bar{B} + \dot{\bar{C}}\,C - \bar{C}\,\dot{C} = 0$
changes under the (anti-)BRST symmetry transformations (47) as:
\[
s_b\,[B + \bar{B} + \dot{\bar{C}}\,C - \bar{C}\,\dot{C}]
 = \Big[\frac{d}{d\,\tau}(B + \bar{B} + \dot{\bar{C}}\,C
 - \bar{C}\,\dot{C})\Big]\,C
-  \Big (B + \bar{B} + \dot{\bar{C}}\,C - \bar{C}\,\dot{C} \Big )\,\dot{C},\]
\[s_{ab}\,[B + \bar{B} + \dot{\bar{C}}\,C - \bar{C}\,\dot{C}] \nonumber\\
= \Big[\frac{d}{d\,\tau}(B + \bar{B} + \dot{\bar{C}}\,C
 - \bar{C}\,\dot{C})\Big]\,\bar{C} 
 -  \Big (B + \bar{B} + \dot{\bar{C}}\,C - \bar{C}\,\dot{C} \Big )\,\dot{\bar{C}}.
\eqno (A.1) \]
 Thus, it can be noted that the (anti-)BRST invariance of the above CF-type restriction is
 valid {\it only} on the submanifold of the space of quantum variables which is defined by the CF-type
 restriction (i.e. $B + \bar{B} + \dot{\bar{C}}\,C - \bar{C}\,\dot{C} = 0$) {\it itself}.
 This establishes the fact that, at the {\it quantum} level, the CF-type restriction is a
 {\it physical} constraint. In fact, our whole {\it quantum} 1D  non-supersymmetric system of the relativistic scalar particle 
 is defined on the submanifold of space of variables where the CF-type restriction is satisfied which,
ultimately, leads to the existence of the coupled (but equivalent) Lagrangians. Furthermore, it is {\it also}
responsible for the absolute anticommutativity ($s_b s_{ab} + s_{ab} s_b = 0$) of the off-shell
nilpotent ($ s_{(a)b}^2 = 0$) (anti-)BRST symmetry transformations $s_{(a)b}$.

The above observation can be captured within the framework of ACSA to BRST formalism,
too. For instance, it can be checked that:
\[
\frac{\partial}{\partial\,\bar{\theta}}\Big[{\tilde{B}}^{(b)}(\tau, \bar{\theta}) + {\tilde{\bar{B}}}^{(b)}(\tau, \bar{\theta}) 
+ {\dot{\tilde{\bar{F}}}}^{(b)}(\tau, \bar{\theta})\,F^{(b)}(\tau, \bar{\theta}) - {\bar{F}}^{(b)}(\tau, \bar{\theta})\,{\dot{F}}^{(b)}(\tau, \bar{\theta})\Big]\]
 \[ = \Big[\frac{d}{d\,\tau}(B + \bar{B} + \dot{\bar{C}}\,C - \bar{C}\,\dot{C})\Big]\,C  - (B + \bar{B} + \dot{\bar{C}}\,C - \bar{C}\,\dot{C})\,\dot{C}
 \equiv s_b\,[B + \bar{B} + \dot{\bar{C}}\,C - \bar{C}\,\dot{C}],\] 
\[\frac{\partial}{\partial\,{\theta}} \Big[{\tilde{B}}^{(ab)}(\tau, {\theta}) + {\tilde{\bar{B}}}^{(ab)}(\tau, {\theta}) 
+ {\dot {\bar{F}}}^{(ab)}(\tau, {\theta})\,F^{(ab)}(\tau, {\theta})  - {\bar{F}}^{(ab)}(\tau, {\theta})\,{\dot{F}}^{(ab)}(\tau, {\theta})\Big]\]
  \[= \Big[\frac{d}{d\,\tau}(B + \bar{B} + \dot{\bar{C}}\,C  - \bar{C}\,\dot{C})\Big]\bar{C} - (B + \bar{B} + \dot{\bar{C}}\,C - \bar{C}\,\dot{C})\,\dot{\bar{C}}
\equiv s_{ab}\,[B + \bar{B} + \dot{\bar{C}}\,C - \bar{C}\,\dot{C}].
\eqno (A.2) \]
 In the above, we have used the following
\[\tilde{B}^{(b)}(\tau, \bar{\theta}) = B(\tau), \quad\qquad \tilde{\bar{B}}^{(ab)}(\tau, \theta) = \bar{B}(\tau), 
\eqno (A.3) \]
due to the observation that $s_b\,B = 0, s_{ab}\,\bar{B} = 0$. It is very interesting to note that the
CF-type restriction: $B + \bar{B} + \dot{\bar{C}}\,C - \bar{C}\,\dot{C} = 0$ is a {\it physical}
constraint on the {\it quantum} theory because it is an (anti-)BRST invariant quantity on a submanifold of the 
space of variables which is defined  by the CF-type equation.\\

 \begin{center}
{\bf Appendix B: On an Alternative  Proof of the Existence of CF-Type Restriction: Anticommutativity of the (Anti-)BRST Charges}\\
 \end{center}
We have provided the proof of the existence of the CF-type restriction in our Sec. 6
by exploiting the virtues of symmetry consideration within the framework of ACSA to BRST formalism. We 
concentrate, in this Appendix, on an {\it alternative}  proof of the existence of CF-type restriction by demanding 
the requirement of absolute anticommutativity of the conserved (anti-)BRST charges which is different
 from our discussions in Sec. 6 where
we have {\it first} imposed the CF-type retraction on the expressions for the conserved 
(anti-)BRST charges [cf. $Q_{(a)b}^{(5)}$] in Eqs. 
(51) and (52) to recast them in the {\it exact} forms w.r.t. BRST and anti-BRST symmetries which have enabled us to prove 
the absolute anticommutativity. To achieve  the above goal in an alternative manner, we {\it directly} apply the BRST symmetry 
transformations $(s_b)$ on the expression for the conserved anti-BRST charge ($Q_{ab}^{(4)}$)  as follows:
\[
 s_b Q_{ab}^{(4)} = e^2\, \Big[\frac {d}{d\tau}\Big\{\Big (\frac {d}{d\tau} \big(B + \bar B 
+ \dot{\bar C}\, C - {\bar C}\, \dot C \big)\Big)\, \bar C\, C +   \, \bar B \,\big(B + \bar B  + \dot{\bar C}\, C - {\bar C}\, \dot C \big)\] 
\[-  \big(B + \bar B + \dot{\bar C}\, C -  {\bar C}\, \dot C \big)\dot{\bar C}\, C \Big\} -   2\, \dot{\bar B}\, \big(B + \bar B +  \dot{\bar C}\, C - {\bar C}\, \dot C \big)\Big]~~~~~~\]
 \[~~~~~~~~~~~~~~ +   2\,e\,\dot e\, C\, \Big[\big(B + \bar B +  \dot{\bar C}\, C - {\bar C}\, \dot C \big)\, \dot {\bar C} -   \Big\{\frac {d}{d\tau}\Big(B + \bar B +  \dot{\bar C}\, C - {\bar C}\, \dot C \Big)\Big\}\, \bar C \Big].
\eqno (B.1) \]
 It is evident, from the above, that {\it every} term on the r.h.s. is {\it zero} provided we impose the (anti-)BRST
 invariant CF-type restriction: $B + \bar B +  \dot{\bar C}\, C - {\bar C}\, \dot C = 0$ from {\it outside}. In other words, 
 the absolute anticommutativity of the (anti-)BRST charges [hidden in the expression $s_b Q_{ab}^{(4)} \equiv -\,i\,\{Q_{ab}^{(4)}, \, Q_b^{(4)}\} = 0$ on
the l.h.s. of (B.1)] is true only in the space of variables where the CF-type restriction is satisfied.

To corroborate the above statement, we now apply the anti-BRST symmetry transformation $(s_{ab})$ on the BRST
charge $Q_{b}^{(4)}$ to obtain the following 
\[
 s_{ab} Q_{b}^{(4)} =  e^2\, \Big[\frac {d}{d\tau}\Big\{\Big (\frac {d}{d\tau} \big(B + \bar B 
+ \dot{\bar C}\, C - {\bar C}\, \dot C \big)\Big)\, \bar C\, C -   B \,\big(B + \bar B  + \dot{\bar C}\, C - {\bar C}\, \dot C \big)\]
\[ -  \,\big(B + \bar B + \dot{\bar C}\, C -  {\bar C}\, \dot C \big){\bar C}\, \dot C \Big\} +   2\, \dot{B}\, \big(B + \bar B +  \dot{\bar C}\, C - {\bar C}\, \dot C \big)\Big]~~~~~\]
\[~~~~~~~~~~~~~~~~ -   2\,e\,\dot e\, \bar C\, \Big[\big(B + \bar B +  \dot{\bar C}\, C - {\bar C}\, \dot C \big)\, \dot {C} -  \Big\{\frac {d}{d\tau}\Big(B + \bar B +  \dot{\bar C}\, C - {\bar C}\, \dot C \Big)\Big\}\, C \Big],
\eqno (B.2) \]
which, once again, demonstrates clearly that the absolute anticommutativity of the (anti-) BRST charges 
is true {\it only} when the entire  theory is considered on a submanifold in the space of {\it quantum} variables where 
the (anti-)BRST invariant CF-type restriction is satisfied.\\

\end{document}